\documentclass[a4paper,fleqn,usenatbib]{mnras}

\usepackage[T1]{fontenc}
\usepackage{ae,aecompl}

\usepackage{graphicx}	
\usepackage{amsmath}	
\usepackage{amssymb}	

\title[Gravitational and mass distribution effects]{Gravitational and mass distribution effects on stationary   superwinds.}
\author[G.A. A\~norve-Zeferino]{G.A. A\~norve-Zeferino$^{}$\thanks{E-mail:
gabriel-alejandro.anorve-zeferino@polytechnique.edu}\\
\'Ecole Polytechnique, Route de Saclay, 91128 Palaiseau, France \\ }
\begin{document}

\date{Accepted date. Received date; in original form  date}

\pagerange{\pageref{firstpage}--\pageref{lastpage}} \pubyear{2010}

\maketitle

\label{firstpage}

\begin{abstract}
Here, we model  the effect of non-uniform  dynamical mass distributions and their associated gravitational fields on the stationary galactic superwind solution.  We do this  by considering  an analogue injection of mass and energy from stellar winds and SNe. We consider both compact dark-matter and baryonic haloes that does not extend further from the galaxies optical radii $R_{\rm opt}$ as well as extended gravitationally-interacting ones. We consider halo profiles that emulate the results of recent cosmological simulations and coincide also with observational estimations from galaxy surveys. This allows to compare the analytical superwind solution with outflows from different kinds of galaxies. We give  analytical formulae that establish  when an outflow is possible  and also  characterize distinct  flow regimes and enrichment scenarios. We also  constraint the parameter space by giving approximate  limits above which  gravitation, self-gravitation and radiative cooling can inhibit the  stationary flow.  We obtain analytical expressions for   the  free  superwind hydrodynamical profiles. We find that the existence or inhibition of the superwind solution highly depends on the steepness  and concentration of the dynamical mass and the  mass and energy injection rates. We  compare our results with  observational data and a recent numerical work. We put our results in the context of the mass-metallicity relationship to discuss observational evidence related to  the  selective loss of metals from the least massive galaxies and also discuss the case of massive galaxies.
\end{abstract}


\section{Introduction}

Powerful outflows of gas are an ubiquitous feature of star-forming galaxies  at both low and high redshift. Early  optical emission-lines  surveys of nearby starburst galaxies  carried out by Lehnert \& Heckman (1995, 1996)  showed that several of the indicators that disclose the presence of a superwind, like  extended line emission, shock-like emission-line ratios and   broad emission lines, were  positively correlated with infrared activity. Nowadays,    the detection of superwinds in nearby galaxies through metal absorption-lines  measurements,  specially  of the Na I D doublet ($\lambda\lambda5890, 5896$), is  an extended  practice (Heckman et al. 2000;  Martin 2005; Rupke \& Veilleux  2005; Rupke, Veilleux \& Sanders 2005a; Martin \& Bouch\'e 2009; and references therein).   Current studies of  blue-shifted absorption lines  have confirmed that many  nearby infrared luminous  ($L_{\rm IR}/L_{\odot}<10^{12}$) and ultraluminous (ULIRGs, $L_{\rm IR}/L_{\odot}>10^{12}$)  star-forming galaxies produce massive superwinds with velocities of several hundreds of km s$^{-1}$. It has been also validated   that the velocity, mass, momentum and energy of these  outflows scale with the galaxy SFR, luminosity and mass (Rupke, Veilleux \& Sanders 2005b);  fact that  seems consistent with the picture of a pressure driven  superwind (Chevalier \& Clegg 1985, CC85 hereafter).  On the other hand, the  bulk properties   of  the  hot X-ray emitting haloes  ($T\sim 10^6$--$10^7$ K) detected around  some  galaxies    (Dahlen,  Weaver \& Heckman 1998; Heckman et al. 2000; Strickland \& Heckman 2009)    also  agree  with the predictions of simple superwind models (see Stevens \& Hartwell 2003).

The spectroscopical evidence of galactic scale outflows at intermediate and high redshifts  is also ample.  Recent infrared and radio studies point out to luminous and ultraluminous infrared galaxies at $z\ge 1$, as the hosts of  the most intense star formation in the universe (P\'erez-Gonz\'alez et al. 2005; Chapman et al. 2005). At redshifts $z=2$--3,  Lyman-break galaxies (LBGs) are probably  the most notable representatives of such extreme behaviour (see  Heckman 2001 and references therein).   The optical and infrared spectra of the LBG population at $z\sim 3$  present    metal absorption lines and   Ly$\alpha$ emission lines  that are respectively blue-shifted and redshifted by hundreds of km s$^{-1}$   with respect to the galaxies rest frames  (Pettini et al.  2001). The same   has been observed  in  optical spectra of lensed Ly$\alpha$ emitting galaxies at $z>4$ (Frye, Broadhurst \& Benitez 2002). The redshifted   Ly$\alpha$ signature has been also detected in   LBGs at $z>5$ (e.g. Dawson, Spinrad, Stern et al. 2002; Tapken  et al. 2007). As it has been  pointed out by Heckman (2001) and Dawson et al. (2002), these observations are also consistent with the picture of an optically  thick  superwind expanding around the star-forming regions.

There is now  a consensus  on that galactic outflows could have a profound impact on the 
 chemical evolution of galaxies and the dynamics of the IGM. They are expected to terminate star formation in some galaxies and to  deposit heavy elements, heat and locally accelerate  the IGM (Nath \& Trentham 1997; Benson \& Madau 2003).  
 
Furthermore,  superwinds seem to be one of the main agents\footnote{Galaxy mergers and tidal effects in dense cluster environments are out of the scope of this work.} in the establishment of  the observed  strong correlation between galaxy mass and metallicity. A substantial amount of observational  evidence points to a selective loss of metals from the least-massive galaxies and a full retention of the same by the most massive ones  (Garnett 2002, Tremonti et al. 2004, Lee et al. 2006).  The studies are coincident  in reporting  a  saturation of the O/H abundance (used as a surrogate for metallicity) for the most massive galaxies and a power-law-like behaviour for the intermediate-mass and least-massive galaxies.  A popular view is that  galactic superwinds are to blame for removing metals from the relative shallower gravitational potential wells of the least-massive galaxies. The absorption-lines and X-ray studies of  superwinds from nearby galaxies carried out by  Heckman et al. (2000)  clearly support this  trend.  However, outliers from this empirical relationship have already been found in the form of low-mass high-metallicity dwarf galaxies (Peeples, Pogge \& Stanek 2008) and massive low-metallicity early-type ones  (Peeples, Pogge \& Stanek 2009).

From another standpoint,  the low metallicity ($Z\sim0.02$--$0.5\, $ Z$_\odot$) and high gas content of many dwarf irregular galaxies  with historials   of ongoing or recently finished starburst  activity  indicate that they are late-type objects. The latter is particularly true for     blue compact dwarf  galaxies (BCDs).  It has been suggested that  these young objects might  be the predecessors  of  the predominantly early-type, gas-poor and low-metallicity population of dE and dSph dwarf galaxies (e.g. Dellenbusch et al. 2008). The favourite theory to explain the gas depletion that such transition  implies is again based on starburst driven superwinds (Larson 1974, Dekel \& Silk 1986, Finlator \& Dav\'e 2008).

The problem of how galaxies retain only certain amount of   metals according to their masses has been already addressed analytically. Lynden-Bell (1992) proposed a simple heuristic model  in which the fraction of the starburst-produced metals that are retained by a galaxy is proportional to the depth of the galaxy potential well for galaxies with escape velocities  less than the outflow effective terminal velocity, i.e.  $v_{\rm e}<V_{\infty}$, and asymptotes to full retention for the most massive galaxies with large  $v_{\rm e}$.   This heuristic approach has been successfully applied by Heckman et al. (2000) and Heckman (2001) to  explain their observational results. Under the assumption of an isothermal gravitational potential,  they proposed a scheme in which an asymptotic full retention of metals  is achieved when  $v_{\rm e}\gg V_{\infty}$ and partial retention is proportional to $v_{\rm e}^2$.

Nevertheless, a self-consistent and simple analytical superwind  hydrodynamical model  incorporating gravitational effects and from which more general conclusions could be reached  is still lacking. CC85 presented the standard galactic superwind model considering just the adiabatic, pressure driven expansion of the hot plasma resulting from the thermalization of individual stellar winds and supernovae ejecta inside of the  starburst volume. Their model applies to fast superwinds for which gravitational effects are weak.  This may be the case for low-mass galaxies; however,  for massive galaxies,  the binding gravitational energy can be comparable to the energy budget provided by the thermalization of the gas injected in the central regions ($v_{\rm e}\sim V_{\rm \infty}\sim1000$ km s$^{-1}$, see Wang 1995). Furthermore, observational studies indicate that in many cases, galactic starburst  episodes are centrally concentrated (e.g. Marlowe  et al. 1995; Taniguchi, Trentham \& Shioya 1998; Cair\'os et al. 2003;  Dellenbusch et al. 2008); an effect that has been so far neglected in superwind analytical models.

Here, we present a simple stationary spherically-symmetric  hydrodynamical model that incorporates  gravitational effects and takes into account a central concentration of the dynamical mass with analogue mass and energy injection rates, with the aim of  addressing, within the natural limitations of our approach, the following issues: (i) How does the galactic gravitational field affect the  superwind hydrodynamical profiles and their related observables? (ii) How does the concentration  of the dynamical mass and the mass and energy injection rates affect the superwind behaviour? (iii) What is the actual value of the asymptotic terminal speed that will determine the impact of the superwind on its surroundings when gravitational fields are taken into account?  (iv) Under what circumstances can the outflow be inhibited? (v) What are the possible enrichment  scenarios and when do they occur?   (vi) What are the implications for the mass-metallicity relationship?  As in previous approaches, a compromise will be established between the two usual suspects of determining the gas fate, $v_{\rm e}$ and $V_{\infty}$ ({see e.g. Sharma and Nath 2012});  however, here it will be done on a purely hydrodynamical basis and  covering  the case of  gravitational potentials that can reproduce asymptotically flat rotation curves. {This latter  fact has not been taken into account in a previous  work by Wang (1995) who presented an analytical superwind model considering power-law and logarithmic gravitational potentials}.

 As it has been previously cautioned, one must distinguish between outflows  localized in extent  (just a few kpc around the star forming region) from those that may be  able to  completely escape  from the galaxies and have an impact on the IGM (see Mac Low \& Ferrera 1999). Here we analyze the former case, since as it has been pointed out by Heckman (2001), the \emph{intrinsic} observable manifestations of galactic superwind are produced by material still relatively deep within the gravitational potential of the galaxy dynamical mass. 
 
An overview of the organisation of the Paper is given next. The superwind model is presented  at the beginning of   Section \ref{model}. It is introduced initially for galaxies with compact haloes which do not exceed the optical radius $R_{\rm opt}$. This can approximate the case of the biggest brightest spirals for which dark matter only accounts 15\% of the total mass (Persic, Salucci \& Stel, 1996) and also SCUBA sources (Silich et al. 2010). In Section \ref{BCs}, the  boundary conditions needed for a supersonic outflow are obtained. We also derive an expression  for the asymptotic terminal speed and provide limits for which gravitation can establish different flow regimes. The associated enrichment scenarios are described qualitatively.   In Section \ref{sol}, we obtain  analytical superwind solutions for the case of compact haloes. In general the dark-matter within $R_{\rm opt}$ can vary from 0\% to 30\%--70\% (Persic \& Salucci, 1997); so,  in Section \ref{Ana_ext_halo} we extend our analytical model to galaxies with extended haloes (i.e. haloes with an extension much larger than $R_{\rm opt}$). An analytical profile for the haloes is specified in Section \ref{halo}. There, we also give the corresponding  limits for the retention and escape of the superwind from its host galaxy.  In Section  \ref{hydrop} we present the resulting hydrodynamical profiles considering superwinds ejected by different kinds of galaxies  and compare with a previous numerical work that considers massive galaxies with uniformly distributed dynamical masses and injection rates (Silich et al. 2010). We contrast the predictions of our model with observational data in Section \ref{MZ}. The conclusions are presented in Section \ref{con}. In Appendixes \ref{gravth} and \ref{coolth},  formulae that establish the circumstances under which self-gravitation and  radiative cooling can inhibit the superwind solution are given.

\section{The superwind model}\label{model}

Let's define first the parameters and variables of the model. The  superwind is powered by a central spheroidal object  that represents either a galaxy or a protogalaxy. Each central object is defined by a set of   \emph{three parameters and a normalised density profile} $\{r_{\rm sc}, \dot E_{\rm eff},\dot M_{\rm eff}, \rho_{\rm  s} \}$:  a characteristic object radius,\footnote{The characteristic radius of a galaxy bulge or nucleus, e.g. the optical radius $R_{\rm opt}$.} $r_{\rm sc}$; the effective energy deposition rate, $\dot E_{\rm eff}$;  the effective mass deposition rate,  $\dot M_{\rm eff}$; and a normalized spatial distribution, $\rho_{\rm  s}$. The latter is used to trace the  densities of the dynamical mass (both the stellar and dark matter components) and the  mass and energy deposition rates inside the galaxy, i.e. they are assumed to be proportional to $\rho_{\rm s}$. {In reality,  the deposition rates are proportional to the star formation rate, which in turn is related to the surface density of gas. So, considering a single $\rho_{\rm s}$ is a  drawback in our model; however, this affects only the central region ($r<r_{sc}$) but gives anyway coherent values of the  hydrodynamical variables at $r_{\rm sc}$}.

 We  have followed Strickland \& Heckman (2009)  in assuming that  $\dot E_{\rm eff}$ and  $\dot M_{\rm eff}$ are given by

\begin{equation}
\dot E_{\rm eff}= \epsilon \zeta \dot E_{\rm SN+SW}
\end{equation}

\noindent and

\begin{equation}
\dot M_{\rm eff} = \zeta \dot M_{\rm SN+SW} + \dot M_{\rm cold}= \beta\zeta  \dot M_{\rm SN+SW},
\end{equation}

\noindent where  $\dot E_{\rm SN+SW}$ and  $\dot M_{\rm SN+SW}$   are the energy and mass deposition rates  due to stellar winds and SNe within the  whole central volume, $\zeta\le1$ is a participation factor that removes  negligible thermalization regions, and $\epsilon \le 1$ is the mean thermalization efficiency. Similarly, $\beta\ge 1$ is a mass loading factor that accounts for the incorporation  of  ambient gas  within the central volume. The \emph{effective} terminal speed is then given by

\begin{equation*}
V_{\infty } = \left( \frac{2\dot E_{\rm eff}}{\dot M_{\rm eff}}\right)^{1/2}=\left( \frac{2\epsilon \dot E_{\rm SN+SW}}{\beta \dot M_{\rm SN+SW}}\right)^{1/2}= 
\end{equation*}

\begin{equation}
12.6 \left( \frac{2\epsilon\dot E_{38}}{\beta\dot M_{\odot}}\right)^{1/2} \, \mbox{km s}^{-1}. \label{Veff}
\end{equation}

\noindent where $\dot E_{38}$ is the energy deposition rate in units of $10^{38}$ erg s$^{-1}$ and $\dot M_\odot$ is the mass deposition rate in $M_\odot$ yr$^{-1}$. The mass, momentum and energy conservation laws  for the flow within $r\le r_{\rm sc}$ are

\begin{equation}
\frac{1}{r^2}\frac{{\rm d} (\rho u r^2) }{{\rm d}r}=\dot q_{\rm m},  \label{CMN}
\end{equation}

\begin{equation}
\rho u \frac{{\rm d}u}{{\rm d}r}=-\frac{{\rm d}P}{{\rm d}r} - \dot q_{\rm m}u-\rho \nabla \phi  \label{CMmN}
\end{equation}
\noindent and

\begin{equation}
\frac{1}{r^2}\frac{{\rm d} \left[\rho u r^2\left(\frac{1}{2}u^2+(\eta +1)\frac{P}{\rho}\right) \right] }{{\rm d}r}=\dot q_{\rm e} -\rho u \nabla \phi. \label{CEN}
\end{equation}

\noindent Here, $r$ is the radial coordinate, $u$ is the velocity, $P$ is the pressure,  $\rho$ is the gas density and $\dot q_{\rm m}$ and $\dot q_{\rm e}$ are the mass and energy deposition rates per unit volume, respectively. Here both are proportional to $\rho_{\rm s}$. We  also have assumed an ideal polytropic flow with   polytropic index $\eta$. The case $\eta=3/2$ is equivalent  to the $\gamma={5}/{3}$ case and $c^2=(\eta+1)P/\eta \rho$ is equivalent to the  squared sound speed (A\~norve-Zeferino, Tenorio-Tagle \& Silich 2009).  The gravitational acceleration  is $-\nabla \phi=-GM(r)/r^2$, where M(r) is the cumulative dynamical mass (see Section \ref{BCs}) and $G$ the constant of universal gravitation.

As it was mentioned in the introduction we concentrate first on the case of compact haloes. So, The conservation laws that govern the flow outside of the central volume ($r>r_{\rm sc}$) are
 
\begin{equation}
\frac{1}{r^2}\frac{{\rm d} (\rho u r^2) }{{\rm d}r}=0, \label{CM}
\end{equation}

\begin{equation}
\rho u \frac{{\rm d}u}{{\rm d}r}=-\frac{{\rm d}P}{{\rm d}r} -\rho \nabla \phi  \label{CMm}
\end{equation}
\noindent and

\begin{equation}
\frac{1}{r^2}\frac{{\rm d} \left[\rho u r^2\left(\frac{1}{2}u^2+(\eta +1)\frac{P}{\rho}\right) \right] }{{\rm d}r}= -\rho u \nabla \phi. \label{CE}
\end{equation}

In the last equation,  $-\nabla\phi=-GM_{\rm DM}/r^2$  and  $M_{\rm DM}$  is the total dynamical mass  of the   central object.   We   neglect the effect of cooling and  self-gravitation.  However, we   evaluate their impact in Appendixes \ref{gravth} and \ref{coolth} providing a justification for this assumption.  Here, we adopt a  symmetry-preserving energy balance approach  --similar to  that present in the CC85 model and in the 3D simulations of Recchi, Matteucci \& D'Ercole (2001)--  that will lead us to relations for comparing the relative strengths of the  thermalization and the gravitational potential in a similar manner than in  the  heuristic proposal  of Lynden-Bell, but with a patent hydrodynamical basis. None the less, in order to compare such strengths, we consider physically motivated mass distributions that can recover flat rotation curves by choosing $\rho_{\rm s}$ in a suitable manner.

We incorporate in our model the effect of  the gravitational field  and obtain an  analytical solution for the external zone ($r>r_{\rm sc}$).  For completeness, we obtain numerically the hydrodynamical profiles corresponding to the inner region ($r\le r_{\rm sc}$). Nevertheless, we use the analytic integrated forms of equations (\ref{CMN}) and (\ref{CEN}) to obtain the proper boundary conditions  at $r_{\rm sc}$  and  unveil some relevant physics, as shown below.

\subsection{Boundary conditions and the existence of the superwind solution} \label{BCs}

We impose boundary conditions that warrant the continuity of the fluxes across the central object surface. The boundary condition for the mass flux, $F_{\rm m}$, is

\begin{equation}
F_{\rm m}(r_{\rm sc})= \frac{\dot M_{\rm eff}}{4\pi} \label{FmBC}.
\end{equation}

\noindent Although trivial, this relation ensures consistency with the stationary mass flux associated to a localised mass injection.

We will focus on superwinds that expand transonically  outside of the central object characteristic radius. Such velocity profiles are only possible if the flow attains a  Mach number equal to unity  at  $r_{\rm sc}$, i.e. if  $u_{\rm sc}=c_{\rm sc}$. In order to  establish the adequate boundary condition for the energy flux, we need to define first  $\rho_{s}$.  We consider  two kinds of  normalized spatial distributions: a truncated version of a profile introduced by Dehnen (1993) and a uniform distribution  (the almost always casted out assumption; see, however, Ji, Wang \& Kwan 2006). Their  respective expressions are
\begin{equation}
\rho_{s}=  \frac{(3-\alpha)}{4\pi}  \left(\frac{r_{\rm sc}+a}{r_{\rm sc}}\right)^{3-\alpha} \frac{a}{r^\alpha(r+a)^{4-\alpha}} \label{Dehnen}
\end{equation}

\noindent and

\begin{equation}
\rho_{s}= \frac{3}{4\pi R_{\rm sc }^3}.\label{homog}
\end{equation}

\begin{figure}
\includegraphics[height=50mm]{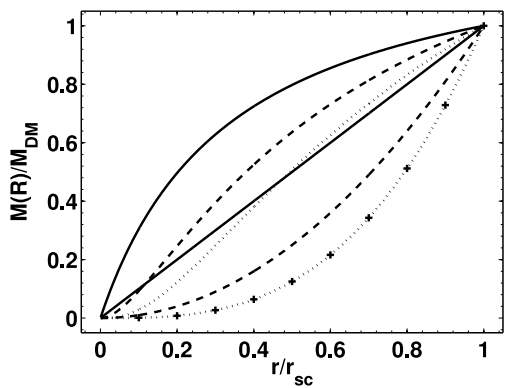}\caption{Normalized cumulative mass for different $\rho_{\rm s}$. The solid, dashed and dotted lines correspond to Jaffe ($\alpha=2$), Hernquist ($\alpha=1$) and plateau-like ($\alpha=0$) truncated profiles, respectively. The concentration parameters are   $A$=0.34  (upper lines) and $A$=1000 (lower lines). The crosses correspond to a uniform mass distribution. }\label{fig1}
\end{figure}

\begin{figure}
\includegraphics[height=50mm]{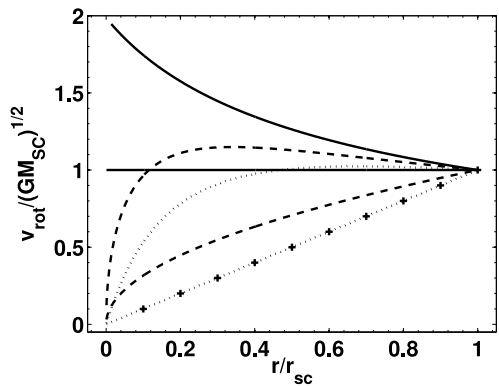}\caption{Rotation speed normalized to  $(GM_{\rm DM})^{1/2}$ for the profiles showed in Fig. \ref{fig1}. Again, the upper lines correspond to $A=0.34$ and the lower lines to $A=1000$. }\label{fig2}
\end{figure}

A huge advantage of the Dehnen-like   distribution is  that it depends on a steepness parameter\footnote{For  reasons related to the convergence  of the energy integral for a spherical stationary flow,  we further constraint the original interval defined by Dehnen: $\alpha\in [0,3).$}  $\alpha\in [0,5/2)$ and a  scale  parameter $a\in (0,\infty)$.   These parameters determine the internal  structure of the central object,  a feature often ignored in analytical superwind models, but that is  decisive in determining the fate of the injected gas.  For  $\alpha=0,1,2$ we obtain  normalized truncated versions of a  plateau-like and the Hernquist (1990) and  Jaffe  (1983) profiles, respectively.  Written in  terms of the normalized radius $R=r/r_{\rm sc}$ and the concentration parameter  $A=a/r_{\rm sc}$, the  explicit expressions for   the cumulative mass corresponding to the  Dehnen-like  profile and the uniform distribution are

\begin{equation}
M(R)=M_{\rm DM} (1+A)^{3-\alpha} \left(\frac{R}{R+A}\right)^{3-\alpha}  \label{Dmass}
\end{equation}

\noindent

\begin{equation}
M(R)=M_{\rm DM}R^3,\label{unimass}
\end{equation}

\noindent  respectively. Equivalent expressions could be written for $\dot E(R)$ and $\dot M(R)$ by replacing $M_{\rm DM}$ by $\dot E_{\rm eff}$ and $\dot M_{\rm eff}$. In Fig. \ref{fig1}, we present the normalized cumulative dynamical mass for  the Jaffe, Hernquist and plateau-like profiles using a small concentration parameter and a  large one. For small values of $A$, most of the mass is contained in the innermost regions of the central object, although at different degrees. For very large values of $A$, the cumulative mass varies almost linearly with $R$ for the Jaffe-like profile; it varies as $\sim R^2$ for the Hernquist-like one and (almost) reproduces the behaviour resulting from the uniform distribution  for the plateau-like profile, i.e.  it varies as $\sim R^3$. For the latter two cases,  most of the mass is contained in the outer layers of the (proto-) galaxy.\footnote{Notice that the behaviour for large $A$  is different from that of the   original Dehnen profiles. This is due to the truncation, since now and respectively,   $\rho_{\rm s}\sim r^{-2} $  $\rho_{\rm s}\sim r^{-1} $ and $\rho_{s}\sim constant$  in the whole volume as $a \rightarrow \infty$.}  The  profiles scale accordingly for other values of $\alpha$ and $A$. The circular velocity profiles, $v_{\rm rot}=\sqrt{GM(r)/r}$,   corresponding to the  distributions showed in Fig. \ref{fig1} are displayed  normalized to $\sqrt{GM_{\rm DM}}$  in Fig. \ref{fig2}.  The escape velocity is given by $v_{\rm e}(r)=\sqrt{2} v_{\rm rot}$. One can recover asymptotically quasi-flat rotation curves for   truncated Dehnen  profiles with $\alpha\le 1$ and small values of $A$. Thus, we can study the effect of the distribution of the dynamical mass and the energy and mass deposition rates per unit volume  in a more general   setting than  previous works. 

With equations (\ref{Dehnen}) and (\ref{homog}), one can explicitly define   $\dot q_{\rm e}=\dot E_{\rm eff} \rho_{\rm s}$, $\dot q_{\rm m}=\dot M_{\rm eff} \rho_{\rm s}$ and $M=4\pi M_{\rm DM} \int_0^r \rho_{s} r'^2 dr'$ and  integrate equations (\ref{CMN}) and (\ref{CEN}) to obtain a Bernoulli-like equation (see A\~norve-Zeferino et al. 2009). From the integration of (\ref{CMN}), equation (\ref{FmBC}) follows immediately.  The integration of (\ref{CEN}) yields 
\[\frac{1}{2}u^2 +(\eta+1)\frac{P}{\rho}=\]
\begin{equation}
\frac{1}{2}V_{\infty}^2 - \frac{1}{(5-2\alpha)} \left(\frac{r_{\rm sc}+a}{r_{\rm sc}}\right)^{3-\alpha} \left(\frac{r}{r+a}\right)^{2-\alpha}\frac{GM_{\rm DM}}{a}\label{I1}
\end{equation}

\noindent and

\begin{equation}
\frac{1}{2}u^2 +(\eta+1)\frac{P}{\rho}=\frac{1}{2}V_{\infty}^2 - \frac{1}{5}\frac{GM_{\rm DM}r^2}{r_{\rm sc}^3}, \label{I2}
\end{equation}

\noindent for the  Dehnen and the uniform distribution, respectively. These   equations will determine the qualitative character of the solution. For instance, note that for the Jaffe-like profile ($\alpha=2$),  the sum of the kinetic energy and the enthalpy per unit mass is reduced everywhere within the central volume by the same amount (i.e. by a constant) due to the steepness of the associated gravitational potential, whereas for the uniform distribution the reduction is  proportional to  $r^2$. 
 
 For convenience and future use we define the dimensionless variable $V_{\rm e}$ as the  squared escape velocity at $r_{\rm sc}$  to the  squared effective terminal speed: 
\begin{equation}
V_{\rm e}=\frac{v_{\rm e}^2}{V_{\infty}^2}= \frac{2GM_{\rm DM}}{r_{sc}V_{\infty}^2}. \label{Ve}
\end{equation}

From the RHSs of equations (\ref{I1}) and (\ref{I2}) one can determine whether  the  inner stationary solution exists or not. There is no solution at all when

\begin{figure}
\includegraphics[height=50mm]{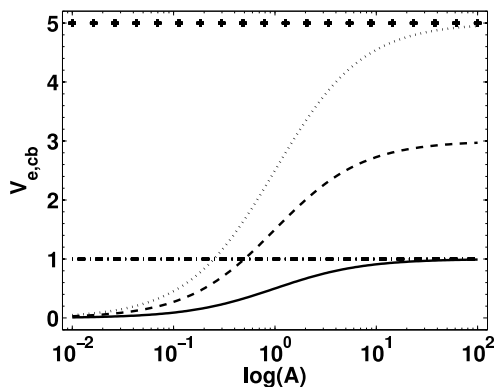}\caption{Threshold lines for the normalized squared escape velocity above which the gas cannot leave the central object, $V_{\rm e,cb}$, as a function of the concentration parameter.   The lines correspond to the same kind of profiles showed in Fig. \ref{fig1}. The dash-dotted line is included as a reference.}\label{fig3}
\end{figure}

\begin{equation}
V_{\rm e}\ge V_{\rm e,cb}= \frac{(5-2\alpha)}{1+\frac{1}{A}} \; \mbox{and} \;   V_{\rm e}\ge  V_{\rm e,cb}= 5, \label{Vecb}
\end{equation}

\noindent respectively. The resulting threshold lines  for $V_{\rm e}$  are shown in Fig. \ref{fig3}.  They are obtained by taking the equality signs  in the equations above; so, according to the case, no solution exists above the respective line. For the Dehnen-like profiles this depends on the numerical values of $\alpha$ and $A$.  

Since $v_{\rm e}^2\propto M_{\rm DM}/r_{\rm sc}$, the more massive and more compact the object, the closer it will be to the respective threshold line and the more difficult  will be for the injected gas to escape from it, as expected. On the other hand, $V_{\rm \infty}^2$ is directly proportional   to $\dot E_{\rm eff}$ and inversely proportional to $\dot M_{\rm eff}$. Hence, poor energy injection rates (low $\dot E_{\rm SN+SW}$), inefficient thermalization ($\epsilon\ll 1$), high mass injections  (large $\dot M_{\rm SN+SW}$) and mass loading ($\beta\gtrsim 1$) can contribute too to maintaining the gas bound, as it is also expected.  However, there are two characteristics of the inequalities in (\ref{Vecb}) that are not so obvious. First, for the uniform distribution the gas can escape from the central object even when $V_{\rm e}>1$. The same occur for models with $\alpha<2$ and intermediate to large values of $A$. Second,  and as it can be seen in  Fig.  \ref{fig3}, the gas can remain bound even when $V_{\rm e}\ll 1$ provided that  $A$ remains sufficiently small. For  models with  $\alpha>2$ the later  can occur for  large values of the concentration parameter $A$.

The first effect occurs because, as explained before,  for the uniform distribution and  the Dehnen  profiles with $\alpha<2$ and large $A$, most of the mass and energy are injected in the outer regions of the (proto-) galaxy, where most of the binding mass (both stellar and dark matter) is also located. A parcel of gas at a position $r$  within this external zone is  driven out   by the  high pressure gradient  more effectively   than it is attracted by   the gravitational force exerted by the relatively voider central regions   (where less mass is concentrated), and thus, the gas   has enough time to cross the object boundary before being pulled back by the gravitational field, although it can be with a rather slow velocity.   The second effect, present  for the Dehnen-like profiles with  relatively small $A$ (this depends on the value of $\alpha$), can be explained in  analogue terms: in this case, the central gravitational  potential is so dominant   that even for small $V_{\rm e}$, the injected gas adopts a steep density profile, and as a result,  the energy per unit mass at $r_{\rm sc}$  --in particular the enthalpy-- is  small. In consequence, the gas cannot escape from the central object. All this implies that only  galaxies  with their parameter $V_{\rm e}$ above the threshold defined by (\ref{Vecb}) would have  deep enough potential wells to keep their injected gas at  $r<r_{\rm sc}$ and enrich  themselves with their processed metals  in a total \emph{closed-box scenario}.

We will proceed to give in advance two results  coming from the analytic solution for the external zone, Section \ref{sol}, equations (\ref{CEi}) and (\ref{CEii}). The first one is that stationary  flows that escape the central region with $V_{\rm e}$ above certain threshold  value will have their stationary solution inhibited in the outer zone, $r>r_{\rm sc}$.  If the stationary central wind has enough ram pressure, it will be difficult for the outer non-stationary flow to cross again the central object boundary, and thus, most likely, a complex morphology, filamentary and/or turbulent (stirred up by galactic rotation),  will result in the outer zone from the interaction of both flows  ('outpouring').   This would be the case when $V_{\rm e}$ is equal or somewhat less than 1. However, the flow can also  be eventually reinserted into the inner zone (Silich  et al. 2010) and rain down over the central object ($1 <V_{\rm e}<V_{\rm e,cb}$, 'inpouring'). The threshold lines  that separate flows in these 'outpouring' (outflow)  and 'inpouring' (inflow/rain back) regimes from flows in a fully stationary regime are given by 

\begin{equation}
V_{\rm e} \ge V_{\rm  e,pour} =  \frac{5-2\alpha}{6-2\alpha+\frac{1}{A}}  \; \mbox{and}  \; V_{\rm e} \ge  V_{\rm  e,pour} = \frac{5}{6}. \label{Vepouring}
\end{equation}

\begin{figure}
\includegraphics[height=50mm]{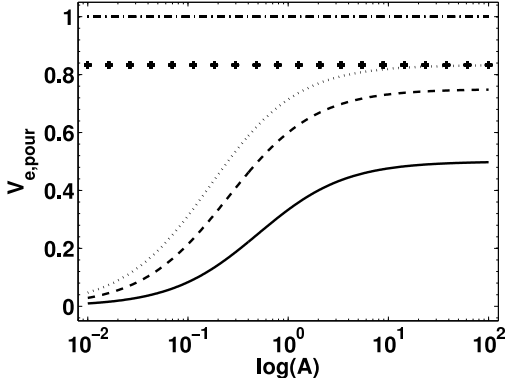}\caption{Threshold lines for the normalized  squared escape velocity  below which the flow is fully stationary.   The lines correspond to the same kind of profiles than in Fig. \ref{fig3}. }\label{fig4}
\end{figure}

\noindent The threshold lines for the cases $\alpha=0,1,2$ and the uniform distribution are presented in Fig. \ref{fig4}. They clearly  reflect  that  for a fully stationary outflow $V_{\rm e}<1$, i.e. the parameters of the flow need to be such that their combination falls below the respective threshold line.

The second result is that not all fully-stationary  superwinds behave in the same way. Some will have bounded (in the mathematical sense) accelerating solutions  and some will have bounded decelerating ones. Using equations  (\ref{CEi}) and (\ref{CEii}),  we find that for a fully-stationary flow,  gravity fixes the \emph{asymptotic  terminal  speed} to\footnote{i.e. the true value of the superwind speed far away from the object, which is clearly distinct from $V_\infty$ defined by equation (\ref{Veff}). Hereafter, we will give  expressions only associated to the truncated Dehnen profiles. The equivalent ones for the uniform  distribution can be obtained from them by taking $\alpha=0$ and $A\rightarrow \infty$.}

\begin{equation}
V_{ \rm g} =\left(1-  \frac{6-2\alpha+\frac{1}{A}}{5-2\alpha} V_{\rm e}\right)^{1/2} V_{\infty}.\label{Vg}
\end{equation}

From   the relation between the  velocity and the sound speed at $r_{\rm sc}$, $u_{\rm sc}=c_{\rm sc}$, it follows that  a fully stationary free superwind  cannot have an accelerating velocity profile if

\begin{equation}
V_{\rm e} \ge V_{\rm e,cons} = \frac{2\eta(5-2\alpha)}{(2\eta+1)(5-2\alpha)+2\eta\left(1+\frac{1}{A}\right)}. \label{Vecons}
\end{equation}

\begin{figure}
\includegraphics[height=50mm]{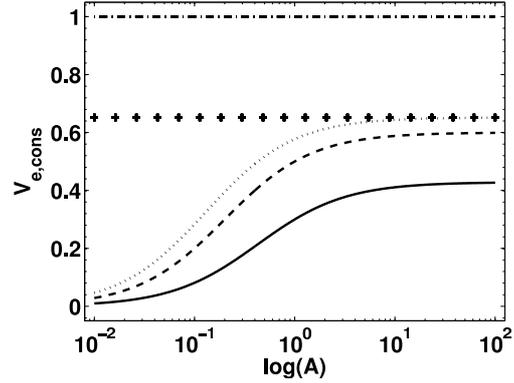}\caption{Threshold lines for the normalized  squared escape velocity  below which the flow has accelerating stationary solutions.   The lines correspond to the same kind of profiles than in Fig. \ref{fig3}. The outflows with decelerating supersonic solutions are compressed between these lines and those shown in Fig. \ref{fig4}. }\label{fig5}
\end{figure}

\noindent In this formula, the equality sign corresponds to an almost  constant external velocity profile with $V_{ \rm g}=u_{\rm sc}$. The corresponding  threshold lines are shown in Fig. \ref{fig5}, where we have assumed  that $\eta=3/2$. This  assumption will be used  in all successive  quantitative calculations.

When the equality in equation (\ref{Vecons}) is satisfied, there is such a concert  between gravity and thermalization that the asymptotic terminal speed depends only on the escape velocity (or rotation speed) at $r_{\rm sc}$:

\begin{equation}
V_{ \rm g,th}=(2\eta)^{-1/2} v_{\rm e}=\eta^{-1/2} v_{\rm rot}. \label{Vgcon}
\end{equation}

\noindent This is a very interesting result; specially, for the case of asymptotically flat rotation curves.

For the particular case $A\rightarrow \infty$ and $\alpha=0$, our analysis concurs with and can explain the numerical results obtained by Silich et al. (2010) for the case of  uniformly distributed dynamical masses and injection rates. They associated the pouring regime with an in-falling  bound wind, which they estimated to occur roughly when $c_{\rm sc}< v_{\rm e}/2$. Such implicit estimation is consistent with  our more exact limit for the  regime. Additionally, they  didn't give any limits for separating bounded accelerating from bounded decelerating fully stationary solutions, although they obtained both types of velocity profiles through their numerical calculations.  Here, we have presented explicit upper and lower analytical  limits for all possible regimes using  more realistic distributions.

Although subtle, the difference between the  lines showed in Fig. \ref{fig4} and Fig. \ref{fig5} is extremely important.  Galaxies with their parameter $V_{\rm e}$  close to the lines showed in Fig. $\ref{fig4}$   will generate (in principle) fully stationary superwinds with very low asymptotic terminal  speeds. Such a  decrease of the velocity can lead to high densities and, in consequence, the flow could  become both gravitationally and radiatively unstable and eventually enter into the outpouring or even the inpouring regime. Since this time,  the gas could have been  polluted by  mixing  with material external to the generating (proto-) galaxy before being reinserted, the inpouring (inflow) regime  corresponds to  an \emph{open-box}  metal enrichment scenario, or perhaps impoverishment or neither of both; see the general theorems presented by Edmunds (1990) and the work of Dalcanton (2007).

Additionally, one must consider that  these  limits  are  general upper bounds. As suggested above, effects like radiative cooling and self-gravitation  can modify the flow.  In Appendixes~\ref{gravth} and \ref{coolth} we evaluate the effect of self-gravitation and also give an approximate analytical expression for the cooling threshold lines.

\section{Analytical  solution for the case of compact haloes} \label{sol}

\subsection{The central superwind}

For the central regions $r<r_{\rm sc}$, we   present a numerical solution to the conservation laws and limit our study to the case $\alpha\le 1$. For such values of  $\alpha$, the  flow have finite central densities   that can withstand self-gravity and catastrophic cooling effects (sections \ref{gravth} and \ref{coolth}) for a wide range of parameters. Thus, for the cases here analyzed,  no bimodal-like behaviour will be obtained when the gravitational field  allows a solution to exist. 

Below, we give the values of the hydrodynamical variables at $r_{\rm sc}$ and the value of the central temperature, which can be useful to characterize the flow:
 
\begin{equation}
u_{\rm sc}^2=   \frac{V_{\infty}^2}{(2\eta+1)} \left(1-\frac{(1+\frac{1}{A})}{(5-2\alpha)} V_{\rm e}\right),
\end{equation}

\begin{equation}
\rho_{\rm sc}= \frac{\dot M_{\rm eff}}{4\pi r_{\rm sc}^2 u_{\rm sc}},
\end{equation}

\begin{equation}
T_{\rm sc}= \frac{\eta \mu_{\rm i}}{(\eta+1)k}  u_{\rm sc}^2, \label{Tsc}
\end{equation}

\noindent and

\begin{equation}
T_{\rm c}= \frac{ \mu_{\rm i}}{2(\eta+1)k}  V_{\infty}^2, \label{Tc}
\end{equation}

\noindent where $\mu_{\rm i}$  is the mean mass per particle for a ionized gas and $k$ the Boltzmann constant.  We remark that  at the threshold for an accelerating solution,  $u_{\rm sc}=V_{ \rm g}=\eta^{-1/2} v_{\rm rot}$.

\subsection{The general solution for the free wind with compact haloes}\label{freewind_compact}

The effect of the gravitational field and the spatial  distribution $\rho_{\rm s}$ on the free superwind hydrodynamical profiles will be  characterized analytically. The   integration of equations (\ref{CM}) and (\ref{CE}) and the application of the boundary conditions (\ref{FmBC}) and (\ref{I1})   yield explicit algebraic relations among the hydrodynamical variables

\begin{equation}
\rho=\frac{\dot M_{\rm eff}}{4\pi u r^2}, \label{CMi}
\end{equation}

\noindent and

\begin{equation}
\frac{1}{2} u^2 + (\eta+1) \frac{P}{\rho}  = \frac{1}{2} V_{\rm g}^2 - \phi,\label{CEi} 
\end{equation}

\noindent where $\phi=-GM_{\rm DM}/r $ and 
\begin{equation}
\frac{1}{2} V_{\rm g}^2 = \frac{1}{2} V_{\infty}^2  -\frac{GM_{\rm DM}}{r_{\rm sc}}-\frac{\left(1+\frac{1}{A}\right)}{(5-2\alpha)}\frac{GM_{\rm DM}}{r_{\rm sc}}.\label{CEii}
\end{equation}

 Again, for convenience, we will  work in terms of   dimensionless variables. For the outer zone they are:

\begin{equation}
R=\frac{r}{r_{\rm sc}}; \;\; U= \frac{u^2}{V_{\rm g} ^2} \;  \mbox{and}\;   \Phi=-\frac{V_{\rm eg}}{R}, \end{equation}

\noindent where $V_{\rm eg}$ is given by

\begin{equation}
V_{\rm eg}= \frac{v_{\rm e}^2}{V_{ \rm g}^2}=\frac{2GM_{\rm DM}}{r_{sc}V_{\rm  g}^2}.
\end{equation}

\noindent  Notice that the normalization of the velocity related terms is now made to the value of the \emph{asymptotic} gravitationally-established terminal speed. After combining equations (\ref{CMi})--(\ref{CEi}) with the equation of   conservation of momentum, we arrive to the following differential equation

\begin{equation}
U'=\frac{\left[1-(U+\Phi)\right]}{(\eta+1)}\left(\frac{2}{R}+\frac{1}{2U}U'\right) + \frac{(U+\Phi)'}{(\eta+1)}-\Phi', \label{EqD}
\end{equation}
 \noindent where the prime symbol indicates differentiation with respect to $R$.
 
 Applying consecutively the changes of variable $\xi= U+\Phi$ and $w=(2\eta+1)\xi-2\eta\Phi-1$ we arrive to an Abel differential equation in non-canonical form. Without loss of generality, let's assume initially that $\eta=3/2$, value that corresponds to a pseudo-adiabatic gas. Thus,  we  obtain
 
\begin{equation}
Rww' = -w^2 + 2 w + 3\left(1+\frac{V_{\rm eg}}{R}\right).
\end{equation}

We can reduce this equation to the canonical form by applying  the change of variable  $W=wR$. The resulting equation is
\begin{equation}
WW'=2W+ 3R\left(1+\frac{V_{\rm eg}}{R}\right). 
\end{equation}

We can further simplify the above  equation if we work in terms of  $x=R+V_{\rm eg}$ as the independent variable and $y=W/x$ as the dependent one. Proceeding this way, the differential equation transforms into a separable one
\begin{equation}
yy'= \frac{- y^2 +2y+3}{x}.
\end{equation}

This last equation is elementary and  can be   integrated using partial fractions. The solution is

\begin{equation}
x= D_{0}(y+1)^{-1/4}(3-y)^{-3/4},\label{x-y}
\end{equation}

\noindent where $D_{0}$ is an integration constant. Returning to our  original dimensionless variables,  we have that 

\begin{equation}
R= D U^{-1/4} \left[ 1-U +\frac{V_{\rm eg}}{R}\right]^{-3/4},
\end{equation} 

\noindent where $D$ is also a constant. By direct substitution, it is easy to show that the generalization

 \begin{equation}
R= D U^{-1/4} \left[ 1-U +\frac{V_{\rm eg}}{R}\right]^{-\eta/2} \label{gensol}
\end{equation} 

\noindent is the general solution of equation (\ref{EqD}) for arbitrary $\eta$.  A similar result  can be obtained for any conservative force   term included in the RHS of the equation of  energy. For our case, the constant $D$ is given by

\begin{equation}
D=\left(\frac{1}{2\eta+1}\right)^{1/4} \left[ \frac{2\eta(1+V_{\rm eg})}{2\eta+1} \right]^{\eta/2}.
\end{equation}

 In the absence of the gravitational field, $V_{\rm eg}$  is identically zero, and thus we recover the CC85 superwind solution written in terms of the velocity (see Cant\'o, Raga \& Rodr\'iguez  2000 and A\~norve-Zeferino et al. 2009).

\subsection{ Branches  and parametric form of the general solution}\label{sec_branches}

Certainly, the inclusion of the gravitational field and the departure from a uniform distribution will modify  the topology of the outer flow.  Nevertheless,  we can use an analogue of  equation  (\ref{x-y})  to gain some insight about the qualitative behaviour of the solution and, for the sake of accuracy and simplicity,  obtain a parametric  form of  (\ref{gensol})  that will avoid  the need of  using a numerical root finder (at least for important values of  $\eta$). For general $\eta$, we have that

\begin{equation}
x(x-V_{\rm eg})^{\frac{2\eta-3}{4}}=  D_0(y+1)^{-\frac{1}{2\eta+1}}(2\eta-y)^{-\frac{2\eta}{2\eta+1}} \label{x-y-eta}
\end{equation}

\noindent with

\begin{equation}
x= R+V_{\rm eg} \label{x},
\end{equation}

\begin{equation}
y=\left[\frac{(2\eta+1)U}{1+\frac{V_{\rm eg}}{R}}-1\right], \label{y}
\end{equation}

\noindent and

\begin{equation}
D_0=  (1+V_{\rm eg})(2\eta)^{\frac{2\eta}{2\eta+1}}.
\end{equation}

\begin{figure}
\hspace{5mm}\includegraphics[height=50mm]{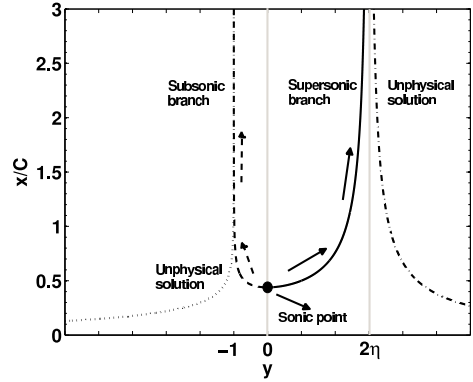}\caption{Branches of the general superwind  solution for $r>r_{\rm sc}$.  For $ y\in (-1,0)$  we have a subsonic expansion. For  $y\in[0,2\eta]$, the expansion is supersonic. } \label{Msol}
\end{figure}

In Fig. \ref{Msol}, we plot    equation (\ref{x-y-eta})  normalized to the value of $D_0$  and  taking the LHS as a function of $y$.  The different branches of the solution are shown.  The values of $y$ corresponding to expanding wind solutions are bounded  to the interval (-1,2$\eta$). Equation (\ref{x-y-eta}) has a single global  minimum at $y=0$ which corresponds to the sonic point.

Hence, The supersonic free wind corresponds to the branch $y\in[0,2\eta)$. By taking $y$ as a parameter, we can obtain the hydrodynamical profiles as follows:

\begin{enumerate}
\item Make  $y$ vary between $0$ and $2\eta$ and  then evaluate the RHS of equation (\ref{x-y-eta}). Then find the respective values of $x$. For $\eta=3/2$ this is straightforward.
\item Using equation (\ref{x}), find the corresponding values of  the normalized radius, $R=x-V_{\rm eg}$.
\item With the values of $R$ and $y$, find U from equation (\ref{y}), $U=(y+1)(1+V_{\rm eg}/R)/(2\eta+1)$. Then de-normalize to find the  actual radius ($r=r_{\rm sc}R$) and   velocity ($u=V_{ \rm g}U^{1/2}$).
\item From equation (\ref{CEi}), the relation $c^2=(\eta+1)P/\eta\rho$ and the equation of state, $P=k\rho T/\mu_{\rm i}$, obtain the rest of the hydrodynamical variables for the (r,u) pairs previously found.
\end{enumerate}

The LHS of   equation (\ref{x-y-eta}) has to satisfy  simultaneously the physical constraint $R\ge 1$ with $R$ a monotonically increasing function of $y$,  and the algebra imposed by the RHS;  however, this not possible for all $r > r_{\rm sc}$  when $V_{\rm e}\ge V_{\rm e, pour}$, and thus, the stationary solution in the outer zone does not exist when the last inequality is satisfied.

\section{The general solution for the free wind on extended haloes} \label{Ana_ext_halo}

Now, we proceed to present the model for an external superwind under the influence of a massive external halo. When external   haloes are included, the equation of conservation of energy outside of the galaxy characteristic radius, equation (9),  transforms into 

\begin{equation}
\frac{1}{r^2}\frac{{\rm d} \left[\rho u r^2\left(\frac{1}{2}u^2+(\eta +1)\frac{P}{\rho}\right) \right] }{{\rm d}r}= -\rho u ( \nabla \phi + \nabla \phi_{\rm h}). \label{CEe}
\end{equation}

Above, the hydrodynamical variables are represented by their usual symbols, $\eta$ is the polytropic index   and  $-\nabla\phi=-GM_{\rm DM}/r^2$,  where  $M_{\rm DM}$  is the total dynamical mass   within $r_{\rm sc}$. Similarly, $-\nabla \phi_{\rm h}=-GM_{\rm h}(r)/r^2$, where $M_{\rm h}(r)$ is the cumulative dynamical mass (i.e. DM+BM) of the external halo, which  has a total mass $M_{\rm H}$.  We will allow the  profile of the external  halo to be defined  either as a continuation of the internal profile or as  a  centrally truncated  profile with different characteristics. 

{When $\nabla\phi=0$ our equations are analogous to the equations of Wang (1995). However, he considered only power-law and logarithmic gravitational potential finding a  solution in terms of the Mach number $M$ as in the CC85 model. Here we consider general halo profiles (i.e. general gravitational potentials) and solve the equations in terms of the velocity giving explicit thresholds for the open-box enrichment  regime (which Wang identified as a "galactic fountain" regime) and the obtention of accelerating superwind solutions.}

The integration of equation (\ref{CEe}) yields a Bernoulli-like equation

\begin{equation}
\frac{1}{2} u^2 + (\eta+1) \frac{P}{\rho}  = \frac{1}{2} V_{\rm g}^2 - \tau \phi - \phi_{\rm h},\label{CEie} 
\end{equation}

\noindent where $\phi=-GM_{\rm DM}/r $,  $\phi_{\rm h}(r)$ is the gravitational potential at $r>r_{\rm sc}$ associated to the non-truncated version of the external halo, $\tau=1-M_{\rm h}(r_{\rm sc})/M_{\rm DM}$ accounts for  truncation effects, and   $V_{\rm g}$  is the asymptotic terminal speed, which is  given this time by 
\begin{equation}
\frac{1}{2} V_{\rm g}^2 = \frac{1}{2} V_{\infty}^2  -\frac{\tau GM_{\rm DM}}{r_{\rm sc}}+  \phi_{\rm h}(r_{\rm sc})-\frac{\left(1+\frac{1}{A}\right)}{(5-2\alpha)}\frac{GM_{\rm DM}}{r_{\rm sc}}.\label{CEiie}
\end{equation}

Note that  $\tau=0$ implies an uninterrupted,  \emph{continuous} gravitational potential.  Similarly, $\tau<0$ implies a centrally truncated external halo with a mass larger than $M_{\rm DM}$, and $0<\tau<1 $ implies the opposite. When $\tau=1$ there is no external halo, and thus $\phi_{\rm h}$ is identically zero. 
 
We will  work again in terms of   dimensionless variables. For the present case they are:

\begin{equation}
R=\frac{r}{r_{\rm sc}}, \; U= \frac{u^2}{V_{\rm g} ^2}, \;   \mbox{and}\;  \Phi=-\tau\frac{V_{\rm eg}}{R} + \Phi_{\rm h} (R);
 \end{equation}

\noindent where $V_{\rm eg}$ is given by

\begin{equation}
V_{\rm eg}= \frac{v_{\rm e}^2}{V_{ \rm g}^2}=\frac{2GM_{\rm DM}}{r_{sc}V_{\rm g}^2},
\end{equation}

\noindent  and $\Phi_{\rm h}(R)$ is $\phi_{\rm h}(r)$ written in terms of $R$ and normalized to $V_{\rm g}^2/2$. The conservation laws can now be reduced to the \emph{same} governing differential equation than in Section \ref{freewind_compact}, see equation (\ref{EqD}).

Thus, within  the theoretical framework developed previously, it is very easy to prove  that the  supersonic free superwind solution  is given by

 \begin{equation}
R= D U^{-1/4} \left[ 1-U +\tau\frac{V_{\rm eg}}{R} -\Phi_{\rm h} (R)\right]^{-\eta/2}, \label{gensole}
\end{equation}

\noindent with

\begin{equation}
D=\left(\frac{1}{2\eta+1}\right)^{1/4} \left\{ \frac{2\eta\left[1+\tau V_{\rm eg} -\Phi_{\rm h}(1)\right]}{2\eta+1} \right\}^{\eta/2}.
\end{equation}

Again, as previously,  we will  give preference to the parametric version of the solution:

\begin{equation}
x[x-\tau V_{\rm eg}+R\Phi_{\rm h}(R)]^{\frac{2\eta-3}{4}}=  D_0(y+1)^{-\frac{1}{2\eta+1}}(2\eta-y)^{-\frac{2\eta}{2\eta+1}}, \label{x-y-etae}
\end{equation}

\noindent where $y$ is a parameter that varies between 0 and $2\eta$ and

\begin{equation}
x= R+\tau V_{\rm eg}  - R\Phi_{\rm h}(R), \label{xe}
\end{equation}

\begin{equation}
y=\left[\frac{(2\eta+1)U}{1+\tau \frac{V_{\rm eg}}{R} -\Phi_{\rm h}(R)}-1\right], \label{ye}
\end{equation}

\noindent and

\begin{equation}
D_0=  [1+\tau V_{\rm eg}-\Phi_{\rm h}(1)](2\eta)^{\frac{2\eta}{2\eta+1}}. \label{D0e}
\end{equation}

To obtain the hydrodynamical profiles, one just needs to follow the algorithm presented  at the end of Section \ref{sec_branches}. An advantage of the parametric solution is that it allows to work with just functions of $R$ in the first two critical steps,  related to equations (\ref{x-y-etae}) and (\ref{xe}). On the other hand, equation (\ref{gensole}) involves both $R$ and $U$. For $\eta=3/2$ (equivalent to the case   $\gamma=5/3$) there is no need for a numerical root finder in the first step of our algorithm. In the second step however,  its use will be most likely unavoidable, as the particular form of the assumed  gravitational potential (i.e. of the external halo profile) is  involved. 

In Section \ref{threse}, we  will give the limit above which the stationary solution is disrupted in the external zone ($r>r_{\rm sc}$) and the necessary condition for  an accelerating stationary superwind solution for the case of extended haloes. In order to do this,  we  will   specify  first the normalized potential $\Phi_{\rm h}$ in the next section.

\section{The extended halo profiles} \label{halo}

How are the DM  and BM  distributed\footnote{we will assume that they together can be specified by a single distribution profile $\rho_{\rm ext}$.} outside of the galaxy characteristic radius? Since we have  permitted   centrally truncated profiles  for the external  halo, theoretically,  we can choose practically any of the usually assumed  distributions; e.g. a NFW profile, Navarro, Frenk \& White (1997); a generalized NFW profile,  Moore et al. (1999);  a self-similar profile, Yoshikawa \& Suto (1999);  an isothermal profile, and so on. Given that      the most commonly used profiles depend  on at least two parameters, and given also the additional freedom introduced by our truncated halo scheme; there is a vast number  of  profiles and  parameters that  can give reasonable agreement with observational studies and with the predictions of cosmological simulations.

We will try to rely on physical insight for  selecting the external halo profile that we will  use in our model.  Recent  cosmological simulations carried out by Abadi et al. (2010) predict that dark matter haloes always contract as a result of galaxy formation. They also found that the contraction  effect is substantially  less pronounced than predicted by the adiabatic contraction model (Blumenthal et al. 1986). On similar grounds, according to  the high-resolution N-body cosmological simulations of  $\Lambda$CDM haloes carried out by  Navarro et al. (2010), the departures from similarity  in the velocity dispersion   and density  profiles   correlate in such a way,  that  a power law for the spherically averaged pseudo-phase-space density is preserved,  $\rho/\sigma^3\propto r^{-1.875}$. They remarked that the  index of the previous power law is identical to  that of a Bertschinger's similarity solution for self-similar infall onto a point mass (in an Einstein-de Sitter Universe). 
They conclude that $\Lambda$CDM  haloes are not strictly universal, but that  the  departure from similarity previously mentioned may be a  fundamental structural property.

Bearing  in mind the results described above,  we conclude that the cases $\tau<0$ and $0<\tau<1$ correspond to   mathematically induced  constraints  that make  continuous the potential at $r=r_{\rm sc}$ for arbitrarily-chosen external-halo  profiles [see equation (\ref{CEie})]. This in turn  might correspond to an external haloe contracted (or expanded) just at the central object edge. The case $\tau=1$  corresponds to the case with no external halo. The case $\tau=0$ is of special interest, as it implies an unforced continuity of the gravitational potential. We will focus on this last case as it turns out that adequately chosen truncated Dehnen profiles satisfy naturally the latter condition.

For  $r<r_{\rm sc}$,  the cumulative dynamical mass  corresponding to a truncated Dehnen profile\footnote{See  also equation (3) in Dehnen (1993).} is given by equation (\ref{Dmass})

\begin{equation*}
M(r) = M_{\rm DM}(1+A)^{3-\alpha}\left(\frac{R}{R+A}\right)^{3-\alpha}.
\end{equation*}

We will also assume a truncated Dehnen profile for the external halo, but we will demand a cumulative mass of the form: 

\begin{equation}
M_{\rm h}(r) = M_{\rm DM}(1+A_1)^{3-\alpha_1}\left(\frac{R}{R+A_1}\right)^{3-\alpha_1}. \label{Moutt}
\end{equation}

 At $R=1$ we have that $M(1)=M_{\rm h}(1)$. Note that for this, we do not require  $A_1=A$ nor $\alpha_1=\alpha$. The last property  can be interpreted in terms  of a contraction of an initial spatial configuration of  DM and BM with concentration $A_1$ and steepness $\alpha_1$ which  produced  a new configuration with concentration $A$ and steepness $\alpha$ for  $r<r_{\rm sc}$, or well, vice-versa, if other processes {{were}} involved (v.gr. angular momentum).   On the other hand, a trivial but  important relationship can be obtained from the condition  $M(1)=M_{\rm h}(1)$ by separating the baryonic and dark matter components:

 \begin{equation}
r_{\rm sc} [M_{\rm bar} + M_{\rm dark}]=r_{\rm sc}  [M_{\rm bar} +M_{\rm dark}]_1.
 \end{equation}
 
\noindent This  could be interpreted as an integral equivalent of the equation for adiabatic collapse derived by Blumenthal et al. (1986). Additionally, given that  the radial velocity dispersion associated to the Dehnen profile goes as $\sigma\sim r^{\alpha/2}$ when $r\rightarrow 0$, we are able to recover the index of the  Bertschinger's  power law near the centre of the galaxy when  $\alpha=3/4$. However,  Navarro et al. (2010) obtained the index from radial averaging, which implies that $\alpha$ can adopt values within a wider range.

Note that  in turn, the previous  configurations could be interpreted as the result of the contraction of an unperturbed configuration   away from the galaxy. This is equivalent to saying that  a galaxy  formed from  the perturbation of an initial state ($A_0$,$\alpha_0$), and that  after certain time, the perturbation bifurcated and produced two inner contracted states characterized by  ($A$,$\alpha$) and ($A_1$,$\alpha_1$). The first state characterizes the inner regions of the  galaxy, $r<r_{\rm sc}$. Then, the characteristic radius $r_{\rm sc}$ can be taken either as   the radius of a     galaxy nucleus or of a bulge. The second state characterizes the outer portions of the galaxy  (e.g. a disc + DM). This is  in agreement with the aforementioned cosmological simulations,  and it implies that galaxies carved out gravitational potential  holes when they formed and that they correspond to local depressions of  an otherwise smoother gravitational potential. 

Here, we are just interested in the superwind solution, so, in order to keep things simple,  we will just consider  the  states $(A,\alpha)$ and $(A_1,\alpha_1)$, i.e.  we will ignore the  depression  of the reference gravitational potential $(A_0,\alpha_0)$. The price that we will pay for this, as well as for the joint distribution of the baryonic and DM components assumed in our scheme, is that instead of (almost) 'perfectly' flat rotation curves  up to 15 times the optical radius (Persic et al. 1996, Salucci \&  Persic 1997), the rotation curves will show some downwards skewness at large radii. They are however very well above the curves corresponding  to   keplerian rotation of  baryonic mass. Evenmore,   the  behaviour of the associated rotation curves  away from $r_{\rm sc}$ is consistent with that of the  universal  rotation curves derived by Salucci et al. (2007) for spiral galaxies. Anyway,  for our purposes, the behaviour at large radii is not that important, as the thermalization driven superwind solution is valid only close to the galaxy\footnote{This implies that the effect of   the 'real' $\Phi_{\rm h}$ can be emulated there by giving adequate values to $A_1$ and $\alpha_1$.}  (see e.g.  Strickland \& Heckman 2009). So, we will proceed to give the expression corresponding to the external gravitational potential.

By taking the limit $R\rightarrow\infty$ in equation (\ref{Moutt}),   one finds that   the  total dynamical mass  is given by  $M_{\rm t}=M_{\rm DM} (1+A_1)^{3-\alpha_1}$.  The expression of the associated gravitational potential for $0\le \alpha\le1$ is then similar  to that given by equation (2) in Dehnen (1993):

\begin{equation}
\Phi_{\rm h}(R)= -\frac{V_{\rm eg}(1+A_1)^{3-\alpha_1}}{(2-\alpha_1) A_1}\left[1-\left(\frac{R}{R+A_1}\right)^{2-\alpha_1}\right].
\end{equation}

With this, we can establish new approximated  thresholds for the open-box enrichment scenario and for accelerating superwind solutions.

\subsection{Thresholds for open-box enrichment and accelerating superwind solutions} \label{threse}

From the energy conservation law, it follows that when the effect of the external halo is considered, the asymptotic terminal speed is given by

\begin{equation}
V_{ \rm g} =\left[1 -\left(\frac{1+\frac{1}{A}}{5-2\alpha}\right)V_{\rm e} + \frac{V_{\rm g}^2}{V_{\infty}^2}\Phi_{\rm h}(1) \right]^{1/2} V_{\infty}. \label{Vge}
\end{equation}

\noindent The  flow enters into  non-stationary  regimes  (inpouring or outpouring)  when 

\begin{equation}
 \left(\frac{1+\frac{1}{A}}{5-2\alpha}\right)V_{\rm e}  - \frac{V_{\rm g}^2}{V_\infty^2}\Phi_{\rm h}(1) \ge 1. \label{outin}
\end{equation}

\noindent  When the above inequality holds,  the galaxy can eventually enter into an open-box enrichment scenario. Otherwise, we will have fully stationary solutions, unless  radiative cooling or self-gravitation inhibit the  stationary solution.

Fully stationary superwinds  have accelerating velocity profiles when

\begin{equation}
-(2\eta+1) \frac{V_{\rm g}^2}{V_\infty^2}\Phi_{\rm h}(1)+2\eta \left(\frac{1+\frac{1}{A}}{5-2\alpha}\right)V_{\rm e}\le 2\eta, \label{accele}
\end{equation}

\noindent otherwise, they have decelerating velocity profiles. When the equality holds in the above relation, we have an almost constant external velocity profile with characteristic velocity
 
\[V_{\rm g}= (2\eta)^{-1/2}[-2\phi_{\rm h}(r_{\rm sc})]^{1/2}= \]
\begin{equation}
\eta^{-1/2}v_{\rm rot}\left\{ \frac{ (1+A_1)^{3-\alpha_1}}{(2-\alpha_1) A_1}\left[1-\left(\frac{1}{1+A_1}\right)^{2-\alpha_1}\right]\right\}^{1/2},
\end{equation}

\noindent where $v_{\rm rot}$ is the rotation speed at $r_{\rm sc}$.

\section{The hydrodynamical profiles}  \label{hydrop}

\subsection{Superwinds on compact haloes}

We will discuss the effect of  concentrated dynamical mass distributions with analogue mass and energy injections  on the hydrodynamics using the  reference models presented in Table \ref{Table1}, which condenses several important cases.

\begin{figure}
	\includegraphics[height=50mm]{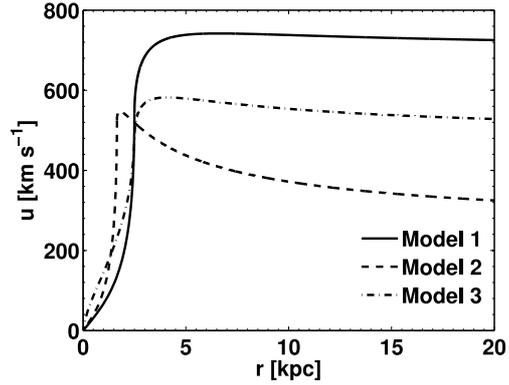}\caption{The velocity profile for the  models representing SCUBA sources.  Here, the profiles for $r>r_{\rm sc}$ were obtained analytically.} \label{fig7}
\end{figure}

{\bf{Models 1 and  2}} correspond to  synthetic  \emph{SCUBA} sources studied by Silich et al. (2010) using their numerical eulerian code, which incorporates both the gravitational field  and radiative cooling. They assumed a uniform distribution of the protogalaxy parameters. They also  considered a continuous star formation scenario and that  mass loading  was proportional to  0.5 times the SFR. According to the Strickland \& Heckman (2009) definition of mass loading, one would require a coefficient $\beta\sim 5.77$ in order to obtain the value of $V_{\infty}$ used by Silich et al. (2010). {\bf{Model 3}} is identical to Model 1, but this time, we have assumed a plateau-like distribution ($\alpha=0$) with concentration parameter $A=0.546$. The corresponding velocity profiles are shown in Fig. \ref{fig7}. There, the profiles for the external zone were obtained analytically.  Because of the effect of the gravitational field, all models have  asymptotic terminal speeds significantly less than $V_{\rm \infty}=1144.6$ km s$^{-1}$. For models 1 and 2, we reproduce the numerical results obtained by the previous authors. Note that  as predicted in Appendix \ref{coolth}, radiative losses are negligible. In addition, as predicted by equation (\ref{Vecons}), Model 1 has a bounded accelerating profile and Model 2 a bounded decelerating one.  Silich et al. (2010) reported a terminal speed of $\sim 740$ km s$^{-1}$ for Model 1 (the value of $u$ at $r\approx 10$ kpc). Using equation (\ref{Vg}), we find that actually, $V_{\rm g}\approx 697$ km s$^{-1}$. Similarly,  $V_{\rm g}\approx 246 $ km s$^{-1}$ for Model 2  and  $V_{\rm g}\approx 479$ km s$^{-1}$  for Model 3. This last model reflects the impact of the   distribution assumed for  the protogalaxy parameters.  We obtain an almost constant velocity profile when the dynamical mass and the energy and mass injections are  more concentrated  towards the object centre, as indicated in Table \ref{Table1}.

{\bf{Model 4}} is based on   parameters fitted to the nucleus of   the  dwarf elliptical galaxy FCC 303  by Hilker et al. (2007, see also Turner et al. 2012). In the optical,  this object  is among  the brightest  in the central region of the Fornax Cluster  and has a central surface brightness profile similar to those of ultra-compact dwarf galaxies (UCDs).  Hilker et al. (2007) used high resolution spectroscopy and surface brightness modeling techniques to derive a  dynamical mass, mass-to-light ratio,   cut-off radius and  half-light radius that are consistent with virial estimators. They concluded that the mass-to-light ratio of FCC 303 is  entirely compatible with a  pure stellar population, i.e. no dark matter is required in order to explain it. {\bf{Model 5}} tries to emulate the most extreme values of the  dynamical mass,  radius and SFR    presented by Peeples et al. (2008) for a sample 43  \emph{isolated} galaxies with morphologies similar to dE and dSph, low-masses  ($M_{\rm DM}\sim 10^{7.4}-10^{10}$ M$_{\odot}$) and  high-oxygen content  ($8.95<12+\log(O/H)<9.3$). The correspondence between our adopted values and those in their sample is not one-to-one though, since we have adopted a  rather small radius for the most massive galaxy. This  was necessary because of  the analogue injection of mass and energy in our hydrodynamical model.  Finally, {\bf{Model 6}} presents the case of an artificial massive  BCD-like galaxy.

For all these models, we have assumed a generic terminal speed of  $V_{\infty} =2500$ km s$^{-1}$ with  $\beta=1$ and $\epsilon=1$. This value fairly agrees  with the superwind recipe given by Strickland \& Heckman (2009) for models that depart from a fixed velocity; however,  here we are considering a different hydrodynamical setting than in their work. So,  afterwards, we adjusted the values of $\beta$ and $\epsilon$ to obtain effective terminal speeds around  $V_{\infty }\sim 1000$  km s$^{-1}$, in order to  get asymptotic terminal speeds that are consistent with the observed typical values (of hundreds of kilometers).   This was done  with the aim of  studying if typical  outflows  can be produced  by  objects with characteristics similar to those present in models 4--6, at a time at   which  they could have been experiencing moderate starburst activity for their type.

\begin{figure*}
	\centering
	\includegraphics[width=50mm] {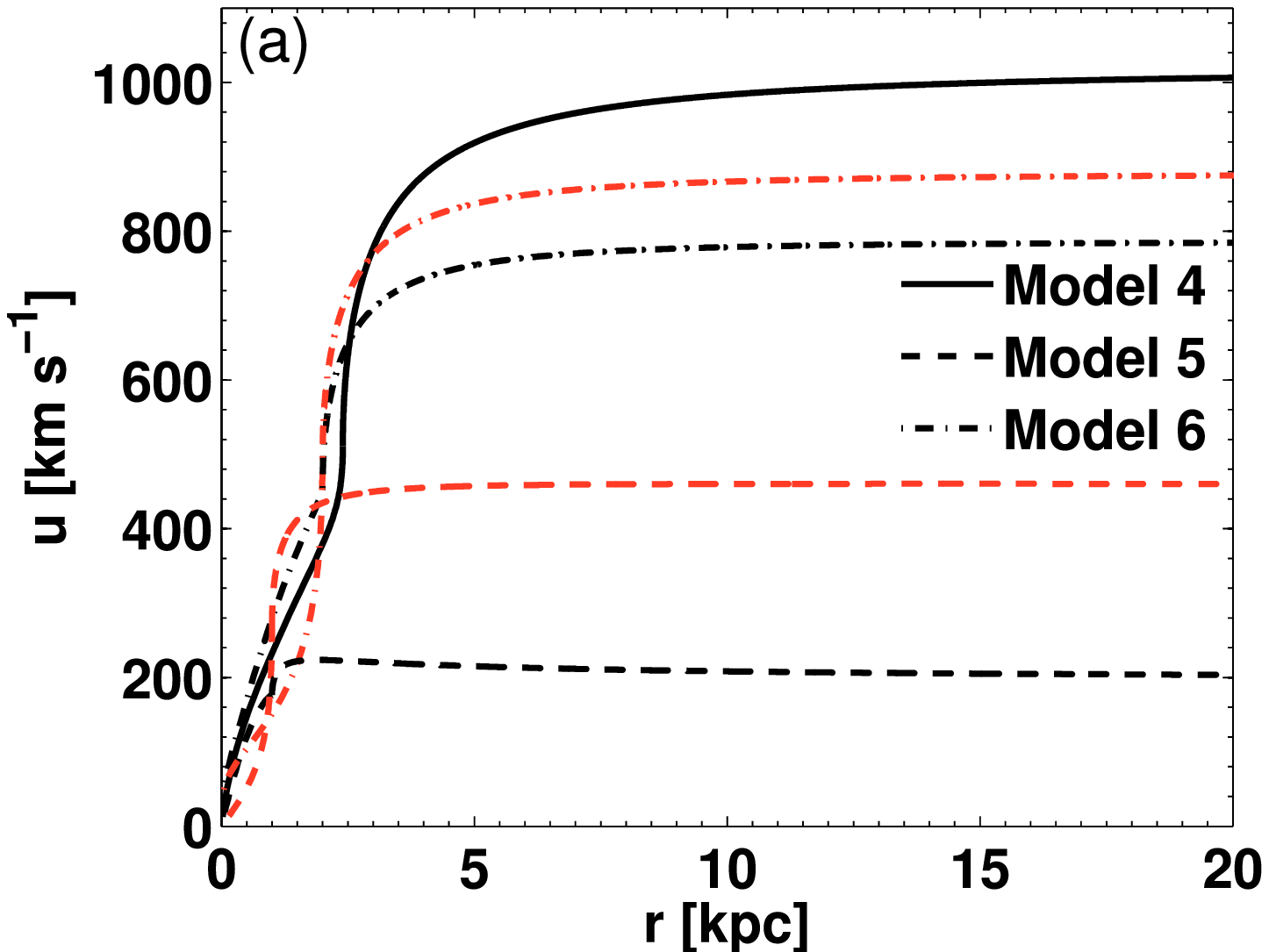}   
	\includegraphics[width=50mm] {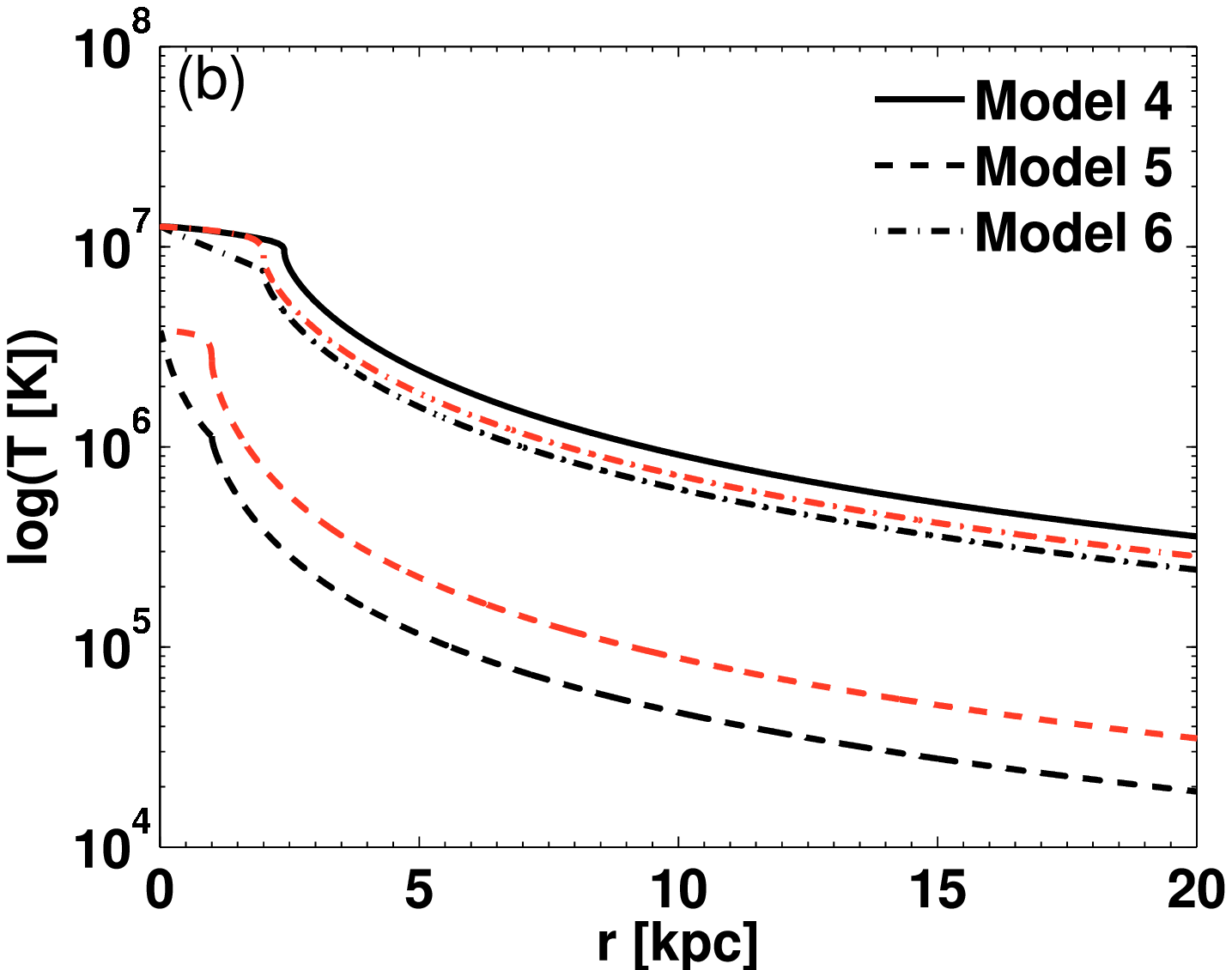} 
	\includegraphics[width=50mm] {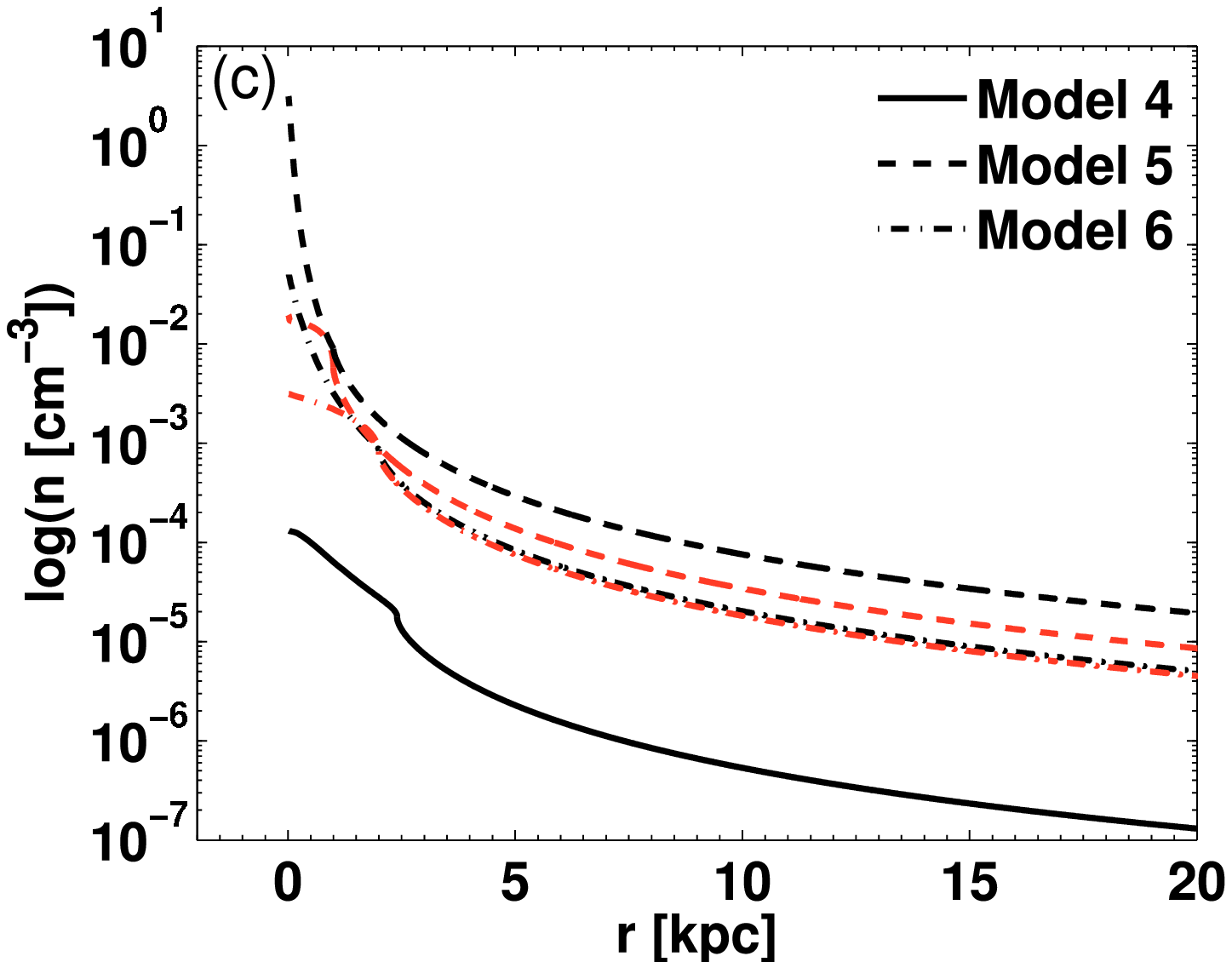} \\ 
	\caption{The velocity, temperature and number density profiles for an object with mass, radius and concentration similar to those of FCC330 (model 4), an isolated dE or dSph galaxy with low-mass and  high concentration parameter (model 5), and an artificial high-mass BCD (model 6). The red lines trace the expected profiles for models 5 and 6 if a uniform distribution is assumed. \label{fig8}}
\end{figure*}

In Fig. \ref{fig8},  we present the velocity, temperature and density profiles corresponding to models 4, 5 and 6. For model 4, which has a moderate concentration parameter,  the external profiles are very similar to the ones predicted by the CC85 model, since they are barely affected by the gravitational field produced by the relatively low mass of the galaxy, in spite of the assumed high mass-loading. As a consequence,   the superwind reaches an asymptotic terminal speed similar to the effective, $V_{\rm g}\sim V_{\infty}$. This is consistent with the view that outflows can more easily remove material from the least-massive galaxies. However,  we found that the central density is larger than the one  predicted by the CC85 model by a factor of $\sim 2$. This implies  that  the  assumed concentration of the starburst   will produce a brighter diffuse emission from the hot gas that remains inside the galaxy.  The internal temperature profile is also somewhat steeper than that predicted by  the CC85 model.

Model 5 is the one that deviates the most from the predictions that can be  extracted from a uniform distribution of the relevant parameters. The asymptotic terminal speed is now  $\sim 0.35V_{\infty}$ instead of  being almost the same. Consequently,  the  temperature of the associated bubble (Weaver et al. 1977) would be reduced by more than $85\%$ and its growing would be compromised.  The characteristics of the  superwind emission are also drastically changed. The central temperature is identical to the one derived  from a uniform distribution, but at intermediate and large radii it is  inferior by a factor of $\sim 2$.  Most remarkably, the central density  is higher by more than two orders of magnitude and the density profile is very steep.  This will translate into much  brighter but compact cores.   At large $r$ the density differs from the predictions of the uniform distribution by a factor $\sim 2$. Model 5 is at the skirts of the limit for   inpouring or outpouring, and very close to  becoming radiatively unstable. A slightly higher concentration  will certainly make  it enter into the open-box or even the closed-box enrichment scenario. On the other hand, a larger SFR would make it enter in a catastrophic cooling regime.

In Model 6, the velocity profile is  flattened by the effect of the large total dynamical  mass and its steeper distribution. The resulting reduction of the asymptotic terminal velocity  with respect to the effective terminal speed,  will   in turn reduce the  post-shock temperature  expected from the standard bubble model by $\sim 40 \%$. The external temperature and density profiles and the internal temperature profile  somewhat differ from the ones that  would result if one assumes a uniform distribution; nevertheless,  drastic differences exist  in the internal  velocity and density profiles. The densities within the innermost regions of the galaxy are  more  than one order of magnitude larger than the values predicted by the uniform distribution. This is produced by a combination of compactness and a steeper dynamical mass profile. This will translate  into an increase of the expected  diffuse X-ray emission (or in other bands provided that the luminosity in band b could be expressed as $L_{\rm b}\propto \int \rho^2 \Lambda_{\rm b}{\rm d}V$) of the central regions   by 1 to 2 orders of magnitude. This can render the  luminosity of the free superwind  at large $r$  dimmer in comparison and also  harder to detect for a given detection threshold. The picture obtained from this synthetic model seems to be consistent with observations of BCDs.

{{\bf Model 7}}  corresponds to  one of the  sets   of parameters found by Strickland \& Heckman (2009)  for M 82 through an extensive observational and theoretical study.  As in their work, model 7 assumes a CC85 model for the hydrodynamics. {{\bf Model 8}}  also corresponds to M 82, but it assumes a steeper distribution ($\alpha=1/2$)  and a concentration parameter that was derived from a  smaller radius also used by   Strickland \& Heckman (2009)  to model this starburst galaxy. As in model 4, given the low mass of the galaxy,  we do not obtain  significant differences between these two models with respect to the CC85 solution, although a steeper distribution can compromise the solution stability.

\begin{table*} 
	\begin{minipage}{175mm}
		\caption{Reference hydrodynamical models}
		\begin{tabular}{@{}llllllllllllll}
			\hline
			Model  &   Type&$\alpha$  &$A$ & $r_{\rm sc}$ & $M_{\rm DM}$ &SFR&$\beta$& $\epsilon$ &  $\zeta$&$V_{\infty}$&  $V_{\rm e}$ &Regime\\
			&&&&(kpc)&($\times 10^{8}$ M$_{\odot}$)&  M$_{\odot}$ yr$^{-1}$& &&& km s$^{-1}$\\
			&&$(\rm a)$&$(\rm b)$&$(\rm c)$&$(\rm d)$&$(\rm e)$&$(\rm f)$&$(\rm g)$&$(\rm h)$&$(\rm i)$&$(\rm j)$&$(\rm k)$\\
			\hline
			1& SCUBA& 0 &$\infty$&  2.5& 2000 &2& $\approx 5.77$& 1& 1& 1144.6&0.5246&accelerating \\
			2& SCUBA&0&$\infty$& 1.65&2000&2&$\approx 5.77$&1&1&1144.6&0.7949&decelerating\\
			3& SCUBA& 0 &   0.5364           &      2.5    &  2000   &   2  &$\approx 5.77$&1&1& 1144.6 & 0.5246 & $\sim $constant, $V_{ \rm g}= u_{\rm sc}$\\
			4&  FCC 303& 0 & 0.6           &2.4   &3.3  &   3  &3&0.5&1&1021&0.0011&accelerating \\
			5&  dE/dSph& 0 & 0.1        &1  &  100   &   0.1   &4&0.2&1& 560&0.2740&borderline\\
			
			6&  BCD& 0.75 & 0.4           &2  &  500   &   3  &3&0.5&1&1021&0.2062&accelerating \\
			7& M 82&0  &    $\infty$         &   0.3      &   7  &  4.5     &1.7&0.55&1&1426&0.0099& accelerating\\
			8&  M 82&0.5  &           0.42   &   0.3      &   7  &  4.5    &1.7&0.55&1&1426&0.0099& accelerating \\

			\hline
		\end{tabular}
		\medskip {\\
			Superwind hydrodynamical models for different kinds of galaxies.  Table headers: (a) steepness parameter, (b) concentration parameter, (c) radius, (d) dynamical mass, (e) star formation rate, (f) mass loading factor, (g) thermalization efficiency, (h) participation factor (i) effective terminal speed, (j) squared ratio of the escape velocity to the effective terminal speed, (k) flow regime.  }  \label{Table1}
		
	\end{minipage}
	
\end{table*}

\subsection{Superwinds on extended haloes} \label{hydroe}

In our model, the  dynamical  mass ($M_{\rm DM}$)  contained within a bulge or galaxy nucleus experiencing an starburst episode  is related to the concentration parameter  and steepness of the external halo  and to  the total dark matter and baryonic mass $M_{\rm t}$:

\begin{equation}
M_{\rm DM}= \frac{M_{\rm t}}{(1+A_1)^{3-\alpha_1}}.
\end{equation}

The dynamical mass contained in the external halo ($r>r_{\rm sc}$) is 

\begin{equation}
M_{\rm H}= M_{\rm t} \left[1 -\frac{1}{(1+A_1)^{3-\alpha_1}}\right].
\end{equation}

Similarly, the  dynamical mass contained up to an external characteristic radius (normalized to $r_{\rm sc}$), $R_{\rm D}$, is
\begin{equation}
M_{\rm D}=M_{\rm t}\left(\frac{R_{\rm D}}{R_{\rm D}+A_1}\right)^{3-\alpha_1}.  \label{Mout}
\end{equation}

The radius $R_{\rm D} $ can be associated to a BM 'disc' radius  or  well  to the BM+DM virial radius. So, all the relevant galaxy parameters  are correlated, in a similar   fashion as in the  work of Salucci et al. (2007).  Nevertheless, we emphasize that the relationship between the parameters is alike but of course not the same, since here we constructed our theoretical model only  following the results of the simulations of Abadi et al. (2010) and Navarro et al. (2010). 

We will proceed to discuss the effect of the extended haloes on the hydrodynamics. In order to do this, we consider the hydrodynamical models  presented in Table \ref{Table1e} and Table \ref{Table2e}.  The first table gives the inner parameters for three galaxies with different characteristics. The second table gives the properties of their external haloes. The groundwork for the discussion will  be  the premise   that the spherical symmetric  superwind solution is a zeroth-order approximation to the aspherical case. We will consider again a reference effective terminal speed of 2500 km s$^{-1}$ for the case of  null mass-loading, fully efficient  thermalization, and total participation within the starburst volume,  i.e. for  $\epsilon=\beta=\zeta=1$.  For  models 2 and 3, the SFRs were obtained from formula (1) in Rupke, Veilleux \& Sanders (2005a) and formula (\ref{Ethcool}), i.e. we  considered SFRs that are  consistent  with the  typical observed luminosities for the object types, and that  in  parameter space, place the objects below the threshold for catastrophic cooling. We find that the predicted temperature profile is barely modified by the presence of the extended haloes. However,  drastic changes are  produced  in the velocity profile.

{\bf{Model 1}} is an extended version of model 5 in Table \ref{Table1},  and corresponds to a synthetic isolated dwarf elliptical  galaxy that tries to emulate the  characteristics of  the most massive outlier of the mass-metallicity relationship detected by Peeples, Pogge \& Stanek (2008). We assumed that the galaxy formed by a contraction of    $\sim 40\%$ of an initially unperturbed   subhalo of DM and BM which had $\sim  70\%$ of its total mass   located within $r\sim 3r_{\rm sc}$, so we used  $A_1=0.5$, $\alpha_1=3/4$ and $R_{\rm D}=1$.  The latter is equivalent to saying that in this case there is no disc, i.e.  we only have a galaxy nucleus.  The internal dynamical mass distribution follows a plateau-like profile, which implies that  some mechanism -- perhaps internal dynamical processes with the action of early powerful superwinds associated to a more extended and powerful starburst episode (see Governato et al. 2010) --  has also transformed the initial mass configuration. In this model,  starburst activity still persists  near the galaxy centre, but with a high concentration.  We assumed a low thermalization efficiency, which implies a small number of massive stars and SNe within the characteristic concentration radius, $A=0.1$. The justification for  this is that the SFR is low,  and that although small, the concentration radius is still much  larger than the typical radius of a massive star, i.e. the filling factor is low. Similarly, because of  the small number of massive stars, just a small incorporation of mass is necessary to produce a heavily  mass-loaded superwind. In this model, the presence of the extended halo suppresses the free superwind solution and the galaxy experiences an open-box enrichment [see equation (\ref{outin})] by keeping the metals processed by the few massive stars  still present near the galaxy centre. This will require however an already gas-poor galaxy  at the moment  at which the pollution occurred       (Peeples, Pogge \& Stanek, 2008).  As suggested above, the required low mass fraction could have been produced by  the action of early  superwinds  associated to  previous and more powerful starburst activity. This is consistent with the views of Peeples et al. (2008), which regarded their sample of outliers as transitional galaxies  in their way to  becoming typically isolated dE and dSph galaxies, but with a high metallicity.  The suppression of the free superwind solution is practically insensitive to {{the}} value of $0\le\alpha_1\le 1$, which indicates that the enrichment is produced by the physical conditions within $A$ and the  initial concentration of the unperturbed subhalo from which the galaxy formed.

\begin{figure}
\includegraphics[height=50mm]{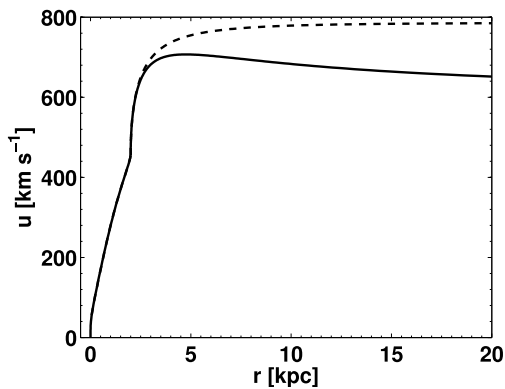}\caption{Superwind velocity profile for model 2 (solid line).  The dashed-line   represents the profile that would result if the external halo were neglected.  } \label{fig1e}
\end{figure}

{{\bf Model 2}} considers the synthetic and very  massive blue compact dwarf  galaxy  modeled previously (see Table \ref{Table1}, Model 6). However, here we add an extended 'disc' to the model in order to 'transform' the galaxy into a  luminous infrared one\footnote{N.B. As LIRGs and ULIRGs,  BCDs may be the result of   mergers, although generally they have lower masses, given that   they mostly form  from the merging  of  dwarf galaxies. Nevertheless, on a higher end, luminous blue compact galaxies  can have dynamical masses of up to $\sim 10^{12}$ M$_\odot$ (Garland et al. 2004, Pisano et al. 2010).} (LIRG, $L_{IR} \sim 10^{11}$ L$_{\odot}$).  LIRGs and ULIRGs may be the end  result of  the merging   of two moderate-size spiral galaxies and display  traces of convergence to an  elliptical  morphology  (Sanders \& Mirabel 1996; Rupke et al.  2005a).  We will model  a LIRG assuming that it  displays a morphology similar to that  of  the central component of Arp 299 (Sargent \& Scoville 1991; Heckman et al. 1999; Hibbard \& Yun, 1999;  Hu et al. 2004), but with just one nucleus. We assume that the disc extends to up to 5 times the radius of the merger nucleus; thus, $R_{\rm D}=5$. The assumed mass and extension are consistent with  CO  emission observations of  (U)LIRGs (Lonsdale, Farrah \& Smith 2006 and references therein). We further assume that the merging process has similarly  transformed the steepnesses of the internal and external mass profiles of the  interacting galaxies unperturbed  haloes,  such that $\alpha=\alpha_1=3/4$.  We adopt the value $A_1=1$ since it produces interesting proportions. In such a case,  $\sim 66\%$ of the BM and DM  of both galaxies is  contained within the warped discs characteristic radius and about $\sim 30\%$ of this fraction resides within the merger nucleus (that is $\sim 20\%$ of the total mass).  As in the original model, starburst activity is present in the nucleus with a somewhat high concentration ($A=0.4$), the thermalization efficiency is 0.5 and mass loading is important, $\beta=3$ (see Heckman et al. 1999). In this model, the gravitational field of the  external halo transforms the accelerating superwind solution associated to the original model into a bounded decelerating one (Fig. \ref{fig1e}). This effect occurs because now we have a more massive galaxy. The produced deceleration will enhance the observable properties of the superwind because of a proportional density increment ($\rho\propto u^{-1}$ ).  However, an even larger total mass   could result in the inhibition of the superwind solution. This is consistent with the superwind  scaling properties found by Rupke, Veilleux \& Sanders  (2005b), whom reported and initial  increment of the superwind observable properties with galaxy mass  and a posterior flattening with the same.

Rupke et al. (2005{{b}}) also reported a flattening of the superwind observable properties at high SFR.  In principle, the normalized free superwind solution is insensitive to the SFR (provided that it could be considered constant during a relatively large time interval), as it just depends on the effective  and asymptotic terminal speeds. However, high SFRs will  intensify the effect of radiative cooling, as more mass will be injected per unit  time and volume, and thus, the stationary solution could also be radiatively inhibited. 
 
\begin{figure}
\includegraphics[height=50mm]{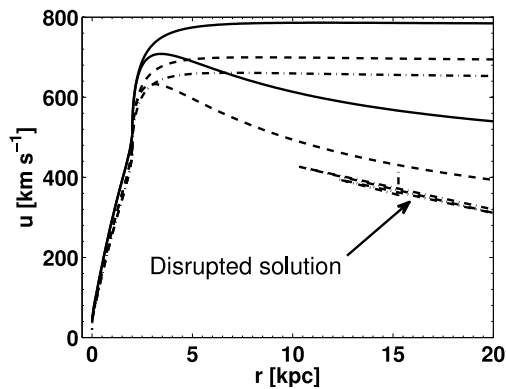}\caption{Velocity profiles for model 3. The solid lines corresponds to the parameters showed in Table \ref{Table1e}.  For this parameters,   $V_{\infty}\approx 1208$   km s$^{-1}$. The lower (upper) solid line  (does not) consider(s) the presence of the external halo.  Similarly, the dashed and dash-dotted lines  correspond to $V_{\infty}=1150$ km s$^{-1}$ and  $V_{\infty}=1125$ km s$^{-1}$, respectively. For the latter case, the stationary free superwind solution does not exist.} \label{fig2e}
\end{figure}

{{\bf Model 3}} gives an extreme example of the effect of the nominal value of the galaxy mass: we model a  massive and  'rare' radio galaxy with a very extended halo (see  e.g. Genzel et al. 2003). We consider a  galaxy with a dynamical  mass of  $1\times 10^{11}$ M$_\odot$ within its nucleus of  $r_{\rm sc}\sim 2$ kpc. A mildly concentrated starburst ($A=0.5$) is present in the nucleus, which has a cuspy dynamical mass distribution ($\alpha=1$).  We consider that the steepnesses of the inner region and the halo are the same and that the total mass of the galaxy is $M_{\rm t}=4\times 10^{11}$ M$_\odot$.  This requires that $A_1=1$. This implies that the half-mass radius is  $r\sim 2.5 r_{\rm sc}$ and that $\sim80 \%$ of the total mass is contained within $r\sim 8r_{\rm sc}$. In this model a high deceleration of the superwind is produced, and    the flow is unstable to small variations of the effective terminal speed (thermalization efficiency), as shown in Fig \ref{fig2e}.  As a consequence, the flow could eventually enter into  the outpouring or even the inpouring regime. On the other hand, if instead of a continuous steepness, we consider that the typical  cuspy halo profile (with slope $\alpha=1$) resulted from the contraction of a smoother one, say with $\alpha_1=3/4$ and $A_1=1$, the  free superwind solution would be inhibited and the galaxy  could enrich itself with is produced metals in an open-box scenario. This would occur because in the second case,  the total mass is slightly larger, $M_{\rm t}=4.75\times 10^{11}$ M$_\odot$. The cumulative dynamical masses of   the two assumed external profiles are very similar, their ratio varies from a value of 1 at $r_{\rm sc}$  (they are identical as they must), up to a value $\sim 0.86$ at $r=10r_{\rm sc}$; nevertheless, such a small variation is enough to suppress the stationary superwind solution. This reflects the fact that at the limit of large galaxy masses, galaxies will retain most of their metals, as expected. 

 \begin{table*} 
\begin{minipage}{175mm}
\caption{Reference hydrodynamical models. Galaxy parameters for $r<r_{\rm sc}$.}
\begin{tabular}{@{}llllllllllllll}
\hline
Model  &   Type&$\alpha$  &$A$ & $r_{\rm sc}$ & $M_{\rm DM}$ &SFR&$\beta$& $\epsilon$ &  $\zeta$&$V_{\infty}$&  $V_{\rm e}$ &Regime (No halo)\\
&&&&(kpc)&($\times 10^{8}$ M$_{\odot}$)&  M$_{\odot}$ yr$^{-1}$& &&& km s$^{-1}$\\
&&$(\rm a)$&$(\rm b)$&$(\rm c)$&$(\rm d)$&$(\rm e)$&$(\rm f)$&$(\rm g)$&$(\rm h)$&$(\rm i)$&$(\rm j)$&$(\rm k)$\\
 \hline
                   1&  dE& 0 & 0.1        &1  &  100   &   0.1   &4&0.2&1& 560&0.2740&borderline\\

          2&  (L)BCD/LIRG& 0.75 & 0.4           &2  &  500   &   $\sim 40$  &3&0.5&1&1021&0.2060&accelerating \\

      3& Radio&1 &   0.5        &      2    &  1000   &   $\sim 200$  &3&0.7&1& 1208& 0.2946 &accelerating\\

\hline
\end{tabular}
 \medskip {\\
 Superwind hydrodynamical models.  Table headers: (a) steepness parameter, (b) concentration parameter, (c) radius, (d) dynamical mass, (e) star formation rate, (f) mass loading factor, (g) thermalization efficiency, (h) participation factor (i) effective terminal speed, (j) squared ratio of the escape velocity to the effective terminal speed, and (k) flow regime when the external halo is neglected.  }  \label{Table1e}
\end{minipage}
\end{table*}

 \begin{table*} 
\begin{minipage}{175mm}
\caption{Reference hydrodynamical models. External halo parameters.}
\begin{tabular}{@{}llllllll}
\hline
Model  &   Type&$\alpha_1$  &$A_1$ & $r_{\rm D}$ & $M_{\rm D}$  &$M_{\rm t}$&Regime\\
&&&&(kpc)&($\times 10^{8}$ M$_{\odot}$)&  ($\times 10^{8}$ M$_{\odot}$)\\
&&$(\rm a)$&$(\rm b)$&$(\rm c)$&$(\rm d)$&$(\rm e)$&$(\rm f)$\\
 \hline

        1&  dE&0.75& 0.5 &1 & $M_{\rm DM}$ & $\sim 2.5 M_{\rm DM}$ & open-box enrichment\\
                      2&  LIRG& 0.75 &1& 5& $\sim 3.16 M_{\rm DM}$  & $\sim  4.76M_{\rm DM}$ & decelerating \\
      3& Radio&1 (3/4)&1&$R_{\rm hm}=2.5$&  $2M_{\rm DM}$ & $4M_{\rm MD}$& decelerating (open-box enrichment)\\

\hline
\end{tabular}
 \medskip {\\
External halo parameters for the models presented in Table \ref{Table1}.  Table headers: (a) steepness parameter, (b) concentration parameter, (c) 'disc' radius (d) 'disc' mass, (e) total mass, and (f) Regime.  In model 3, $R_{\rm hm}$  corresponds to the half-mass radius. }  \label{Table2e}
\end{minipage}
\end{table*}

We next present a comparison with observational data  and a discussion of the implications of these results for the mass-metallicity relationship.

\section{Discussion and comparison with observational data}  \label{MZ}
\begin{figure*}
\includegraphics[height=55mm]{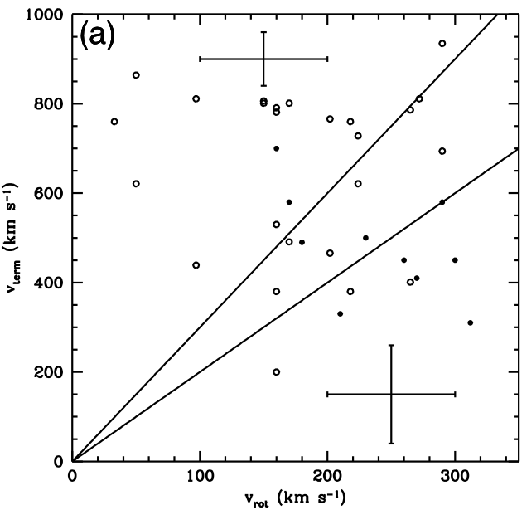}
\vspace{6mm}
\includegraphics[height=55mm] {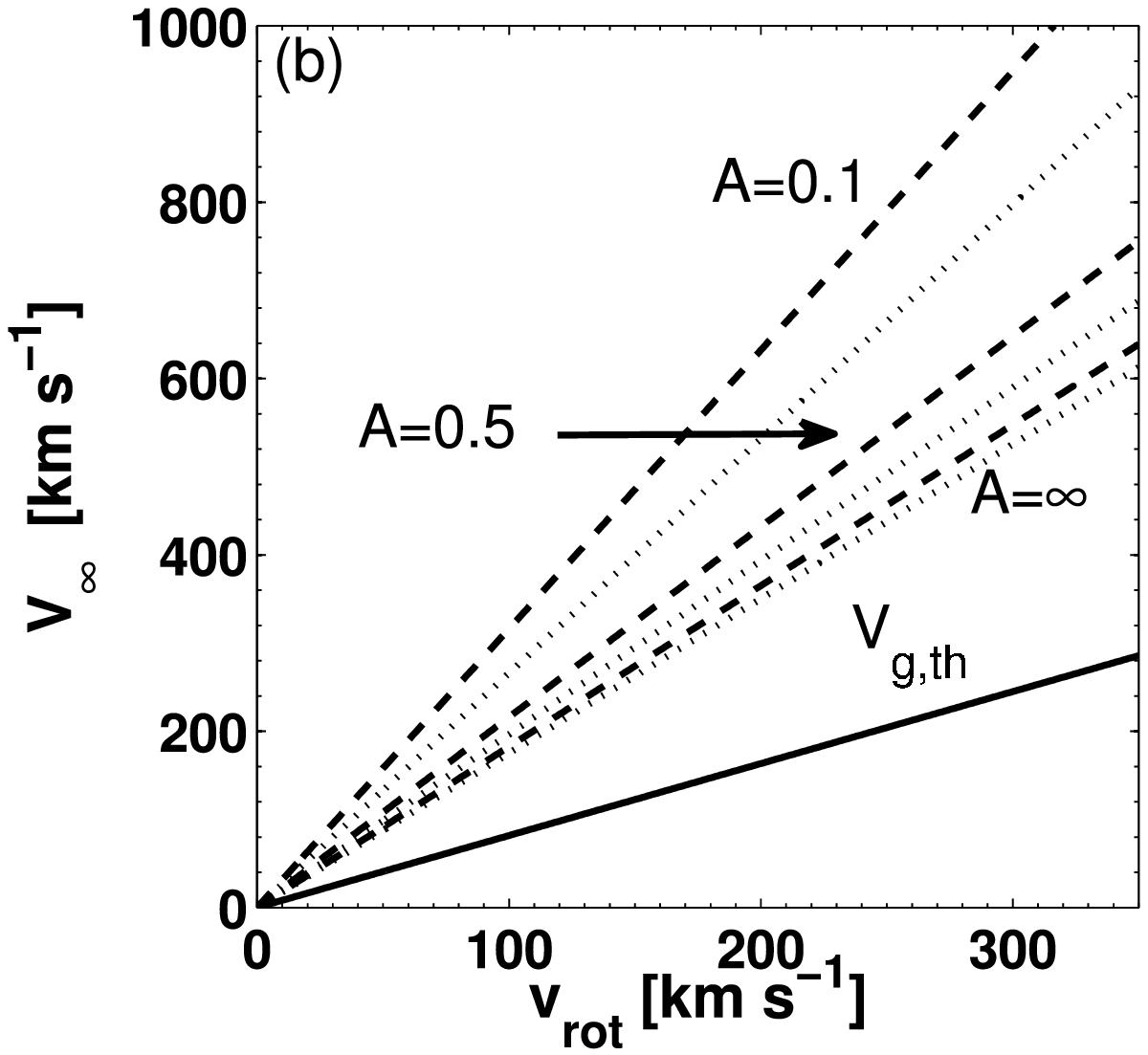} 

\caption{Panel (a): fig. 12 from Heckman et al. (2000), reproduced by permission of the AAS.  The figure shows the galaxy rotation speed vs. the inferred terminal speeds of  superwinds produced by nearby galaxies. The terminal speeds were derived by  Heckman et al. (2000) from   X-ray  (hollow dots) and  Na D  (solid dots) observations. The two diagonal lines indicate the galaxy escape velocity under the assumption that $v_{\rm e}=2v_{\rm rot}$ and $v_{\rm e}=3v_{\rm rot}$. The lower (upper) error bar indicates the typical uncertainties in the X-ray (Na D) estimations of $V_\infty$. Panel (b): threshold lines for the effective terminal speed needed to establish an accelerating superwind solution as a function of the galaxy rotation speed. The assumed concentration parameters are indicated in the figure. The dashed lines correspond to $\alpha=1$ and the dotted lines to $\alpha=0$. The solid line indicates the minimum asymptotic terminal speed that an accelerating outflow must reach for a given rotation speed. }  \label{fig9}
\end{figure*}

In panel (a) of Fig.  \ref{fig9}, we present the effective terminal speeds derived by Heckman et al. (2000) from X-rays and Na D absorption-lines observations of superwinds in nearby  galaxies.  As a reference, these authors included the lines $v_{\rm e}=2v_{\rm rot }$ and  $v_{\rm e}=3v_{\rm rot }$. Their data suggest that superwinds escape only from galaxies with small rotation speeds, which in turn indicates shallow gravitational potentials and low masses. In panel (b) of the same figure, we present  the lines above which accelerating superwind  solutions can be obtained for different concentration parameters and steepnesses. This lines come from equation\footnote{$V_{\infty }^2=2v_{\rm rot}^2/V_{\rm e,cons}$.} (\ref{Vecons}), and we have assumed  a plateau-like and a  Herquist-like profile. The latter reproduce the  behaviour of cuspy dark matter haloes. The former is more adequate to model the matter distribution of  dwarf galaxies. The dashed and dotted lines represent  thresholds for the  {effective} terminal speed. Thermalization and mass-loading inside of a (proto-) galaxy have to occur in such a way that $V_{\infty}$ has to be larger than the respective threshold value  if  an accelerating superwind is to be produced.

As it is shown in Fig.  \ref{fig9}, the higher the  concentration  (i.e. for smaller  $A$) the more stringent are the requirements for producing an outflow, as lower mass injections,  poor mass-loading  and  a higher and more efficient energy injection inside of galaxies would be  necessary in order  to reach the  needed value of $V_{\infty}$. At first sight, the sample of Heckman et al. (2000) seems to  correspond to galaxies with intermediate concentrations, given that the line $v_{\rm e}=2v_{\rm rot}$ in their figure is similar to the lines for $A=0.5$. However, some caution needs to be exerted. As  Heckman et al. (2000) have indicated, for the X-rays data,  they adopted a conservative approach in deriving the effective terminal speed by associating  the observed X-ray temperature  to the central temperature predicted by the CC85 model. In our model, the X-ray temperature is between the value derived by Heckman et al. (2000), equation (\ref{Tc}), and a smaller value  that depends on the intensity of the  gravitational field and on the kinetic energy, equation  (\ref{Tsc}). This effect can be relevant  for massive galaxies, specially given that  projection effects will in turn determine the observed value of the X-ray temperature (A\~norve-Zeferino et al. 2009).  However, when pertinent, the consideration of such effects would only displace the X-ray data in  Fig.  \ref{fig9} upwards.  This will reflect the fact  that deeper gravitational potentials also impose  a more stringent condition over the required  value of $V_{\infty}$ and will provide additional support for the Heckman et al. (2000) conclusion that superwinds can remove  metals more easily from the least-massive galaxies.

On the other hand,  the Na D data is more suitable for representing the  {asymptotic} terminal speed. In Fig. \ref{fig9}b we also display  the expected value of its  threshold, $V_{ \rm g,th}$, equation (\ref{Vgcon}). One must remind that  such threshold  depends just on $v_{\rm rot}$.  Both the Na D  and the  X-rays  data are completely above this limit. This is exactly what should be expected according to our model and the observed lack of correlation between the velocity dispersion in the absorbing material with the galaxies  rotation speed (Heckman et al. 2000): if  the effective terminal speed satisfies the requirements imposed by gravity and the concentration of the starburst for an accelerating solution ($V_{\rm e}<V_{\rm e, cons}$), the resulting superwind  will reach  an asymptotic terminal speed $V_{\rm g}>V_{\rm g,th}$ and no correlation with the galaxy rotation speed will be observed. However,  one could have expected that  departures from spherical symmetry, local effects, the presence of a disc, and all the usual  'buts' would have produced at least some deviations. It seems  withal that the threshold  $V_{\rm g,th}$ is a reliable lower limit.

Then again, the difficulties imposed by highly concentrated starburst for outflows can be an argument for explaining the high metallicity measured by Peeples et al. (2008) for their sample of low-mass  outliers of the M-Z relationship. The galaxies in their  sample do not occupy  unexpected positions in the color-magnitude diagram, have normal SFRs and  are not unusually compact.  They point out that the only remarkable morphological characteristic  in their sample is the presence of bright and often very blue compact cores in ten of the galaxies. They also  suggest  that their sample represents transitional  dwarf galaxies at the end of their star formation activity and at the edge of becoming typical isolated dE and dSph galaxies. However, their high metallicities do not correspond to the expectations for BCDs.  Since the galaxies  they selected are isolated and non-interacting, environmental effects can be discarded (Ellison et al. 2009). In terms of a closed-box enrichment model, Peeples et al. (2008) explain that  the only possibility  is that these galaxies might have  low gas fractions  for their masses. In such a case, only a small pollution would be enough  to enrich the  gas  (Dalcanton 2007).   On  similar grounds,  Ellison et al. (2008) analyzed a large galaxy sample from SDSS and determined  that at fixed mass, galaxies with smaller half-light radii tend to have higher abundances. They proposed that superwinds could be responsible of the selective loss of metals.

We also suggest that the above effects  are or were produced by superwinds struggling against  the  sharp central gravitational potential  of the concentrated starburst episodes, which might correspond to blue cores.  The model here presented (see Model 1 in Section \ref{hydroe}) can  provide a basis to  qualitatively   explain  these results, as well as to quantitatively evaluate if the superwind hypothesis  can approximate the observed dispersion of the mass-metallicity relationship (Tremonti et al. 2004).

Finally, we remark that our model has the scaling properties reported  by Rupke et al. (2005b).  It  has inherited the scaling  with star formation rate  and effective terminal speed  of the CC85 model. On the other hand, its scaling properties with respect to galaxy mass  and radius are different because of the incorporation of the gravitational field.  For a fixed SFR, Rupke et al. (2005b)  reported the enhancement of the superwind properties with increasing galaxy mass  as an unexpected result. We have shown that at fixed SFR and starburst concentration,  this  can be explained in terms of  the larger densities of the outflows associated to the most massive galaxies; so, the larger the galaxy mass the  more intense the observable  manifestations of the superwind. However, this cannot continue indefinitely, as equations   (\ref{Vepouring}) and (\ref{outin}) establish   limits above which  galactic superwinds   eventually enter into the inpouring or outpouring regimes. In turn,  this can explain the observed flattening of the observable superwind properties for large galaxy masses and SFRs also reported by Rupke et al. (2005b). As an alternative to explain the flattening, those authors   proposed heuristically the existence of a terminal velocity for ULIRGs \emph{above} which superwinds cannot be accelerated, and/or  a reduction in thermalization efficiency at high SFR.  According to our model, we find the opposite: there exist an asymptotic effective terminal velocity \emph{below} which no accelerating superwind exists and such velocity depends on   the galaxies rotation speed (or mass and radius), concentration  of their starburst episode, and the steepness of the distribution of their dynamical mass, see equations (\ref{Vecons}) and (\ref{accele}).

\section{Conclusions}  \label{con}
In this theoretical work we presented an analytical model that permits to predict the impact of the the gravitational field on the free superwind stationary solution. Our general findings are:

\begin{enumerate}
\item The existence or inhibition of the stationary superwind solution highly depends on the concentration and steepness of the dynamical mass and mass and energy injection rates. A superwind can be more easily  inhibited when the steepness and concentration are high.
\item We found that the gravitational field fixes the asymptotic terminal speed that will determine the impact of the superwind in the surroundings, equations (\ref{Vg}) and (\ref{Vge}).
\item The gravitational field can establish different flow regimes and enrichments scenarios that also depend on the steepness and concentration of the galactic parameters: close-box enrichment, equation (\ref{Vecb}); either non-stationary outflows (outpouring) or  open-box enrichment/impoverishment (inpouring), equations  (\ref{Vepouring}) and (\ref{outin}); and either  accelerating  or decelerating  stationary outflows, equations (\ref{Vecons}) and (\ref{accele}).
\item We have established the limits above which self-gravitation and radiative cooling can inhibit the stationary solution. Self-gravitation is likely to be unimportant effect in most cases. Cooling on the other hand affects  preferentially to the most massive and compact galaxies with  highly concentrated masses and injection rates.
\item We have evaluated the impact of the gravitational field on the hydrodynamical profiles. We find that the gravitational field can drastically change the expected expansion rate and X-ray emission from the superwind.  
\item  We find that the gravitational field of the extended haloes associated to massive galaxies can drastically alter the free superwind velocity profile and enhance its observable properties. We also find that massive haloes  can also contribute to the inhibition of the superwind solution. 
\item Our model can explain both  the observed initial enhancement and  posterior flattening  of the superwind  properties with the galaxy parameters.
\item Our model is in agreement with observational data that support the view that metals selectively escape from the least-massive galaxies. However, we demonstrated that, under certain circumstances, a high concentration (i.e. a small concentration parameter $A$) can change this. 

\item In our model, the galaxy total mass (BM+DM), the mass contained within a  bulge or galaxy  nucleus (defined by the characteristic radius $r_{\rm sc}$), the mass up to the disc characteristic radius,  and the steepness and concentration of the external halo, are all correlated.  Since  the correlations  are nonlinear,  deviations from galaxy to galaxy are permitted, see Tables \ref{Table1e} and \ref{Table2e}.  Oppositely,  we assumed no correlation between  the above parameters and the concentration and steepness of the mass distribution for $r<r_{\rm sc}$. This is consistent with the results of the cosmological simulations carried out by  Abadi et al. (2010) Navarro et al (2010), in the sense that haloes are not strictly universal. This should be expected, as we based our model in the 'structural contraction' property  derived from their simulations. On the other hand, in their  extensive work, Salucci et al. (2007) found   that the previous parameters were correlated for spiral galaxies, and proposed universal rotation curves assuming  a  Burkert  (1995) profile  for the DM distribution. Our theoretical work diverges from theirs in that we considered additionally the mentioned inner concentration and steepness,  which traces starburst episodes. Such a consideration discards the possibility of universal halo profiles and rotation curves, since in general this parameters will differ from galaxy to galaxy ($r<r_{\rm sc}$); however, the discrepancy smooths out at larger radii, and thus one could talk of an  'asymptotically universal' property, in the sense defined by Salucci et al. (2007).

 \end{enumerate}
\section{Acknowledgements}  \label{Ack}

We   thank to  the {AAS}  and Timothy Heckman for kindly  granting permission for reproducing here  fig. 12 from Heckman et al. (2000). We are also grateful to Timothy Heckman for   answering our  questions about his data.

\appendix

\section{Self-gravitation} \label{gravth}

For a continuous, non-decelerating solution the density  and the sound speed are  monotonically decreasing functions of $r$. Thus, self-gravitation becomes important at any radius at  which  the gas density becomes comparable to the Jeans density.  If the superwind is able to escape from a (proto-) galaxy, the density at $r_{\rm sc}$ has to be less than $\rho_{\rm J}=c_{\rm sc}^2/Gr_{\rm sc}^2$. Using equations (\ref{CMi}), (\ref{I1}), and (\ref{I2}) together with the condition $u_{\rm sc}=c_{\rm sc}$  we obtain that no solution exists  when

\begin{equation}
V_{\rm e} \ge  V_{\rm e,sg} = \frac{(5-2\alpha)}{\left(1+\frac{1}{A}\right)} \left[1-\left(
\frac{4(2\eta+1)^3G^2\dot E_{\rm eff}^2}{(4\pi)^2V_{\infty}^{10}} \right)^{1/3} \right].  \label{VeSG}
\end{equation}

The expression between square brackets further reduces the permitted values of $V_{\rm e}$. An  abrupt cut-off occurs when the quantity between parenthesis inside squared brackets becomes unity. We can rewrite this quantity  as   $[R^4/u_{\rm sc}^4](G^2 \rho_{sc}^2) =t_{\rm dyn}^4/t_{\rm ff}^4$, i.e. as the fourth power of the ratio of the  dynamical time to the free-fall time. Hence, it recovers the natural time-scales to evaluate self-gravitation effects. When $V_{\rm e}$ satisfies  inequality  (\ref{VeSG}), both the  gravitational field associated to the dynamical mass and self-gravitation inhibit the solution. When  $t_{\rm dyn}^4/t_{\rm ff}^4=1$, the responsible is pure self-gravitation. The latter,  however, would require extraordinary large energy injection rates and/or mass-loading, see equation (\ref{Veff}).  Hence, self-gravitation will be an unimportant effect in most cases.

However,  for some of the Dehnen-like distributions, self-gravitation (and radiative cooling)  can  inhibit the stationary  flow  for  $r\ll r_{\rm sc}$.  This is always the case for the central regions of (proto-) galaxies  with density distributions steeper than the Herquist-like profile\footnote{v.gr.  those  that follow the Jaffe-like profile, which nevertheless, reproduces the de Vaucouleurs law and can have either an absolute flat or a quasi-flat $v_{\rm rot}$ curve with a central cusp.} ($\alpha>1$), regardless   of their  other parameters.  This is a consequence of the conservation of mass.  For such profiles, the injection of mass per unit volume  near the object centre  is so intense that the resulting gas density would be extremely high (mathematically, infinite). Hence, self-gravitation and cooling will be important there and, as a consequence, the stationary solution will not be continuous in the whole central volume. Instead,  it  will  adopt bimodal solutions  similar to the radiatively induced ones presented by Silich et al. (2010), in which a stationary solution exist only for $r$ larger than a stagnation radius, $r_{\rm st}$.  Here, we will limit ourselves to the case $\alpha\le 1$ with parameters located below the threshold imposed by the Jeans criteria and our radiative cooling threshold lines, which we present below.

\section{Radiative cooling} \label{coolth}

We use a weighed leading  order approach to assess the impact of radiative cooling   for the case   $0\le \alpha\le 1$, i.e. we  first expand the velocity in a  series of  the form 

\begin{equation}
u\sim \frac{ r^{-\alpha}}{\xi}\sum_{k=1}^{\infty} d_{k}r^k
\end{equation}

\noindent  and take just  the leading order term, such that $u\sim  d_{1} r^{1-\alpha}/\xi$.  Here, $\xi$ is the weight factor.

We then integrate equation (\ref{CMN}) to obtain

\begin{equation}
\rho = \frac{\dot M_{\rm eff}}{4\pi r_{\rm sc}^2} \left(\frac{1+A}{R+A}\right)^{3-\alpha}\frac{R^{1-\alpha}}{u}. \label{rhoc}
\end{equation}

\noindent Using this equation and the approximation for   $u$,  we  get an equivalent of equation (\ref{I1})  for the radiative case.  Written in  terms  of  dimensionless variables, such equation is

\[
U+2 \eta C =1 -\frac{1}{(5-2\alpha)}\left[\left({1+A} \right)^{3-\alpha} \left(\frac{R}{R+A}\right)^{2-\alpha} \frac{V_{\rm e}}{A}\right.
\]

\begin{equation}
\left. - \frac{4 \dot E \Lambda\xi^2  f(A,\alpha,R)}{4\pi\mu_{\rm n}^2 r_{\rm sc}V_{\infty}^6U_{\rm sc}A} (1+A)^{(3-\alpha)} R^{3-\alpha} \left(\frac{1}{R+A}\right)^{2-\alpha}  \right],
\end{equation}
\noindent where $U$ and $C$ are $u^2$  and $c^2$ normalized to $V_{\infty}^2$,  $f$ is a function that $ \rightarrow (5-2\alpha)/3$ as $A\rightarrow \infty$ and blows up  at $R=1$ for  really  small values of $A$ and $\alpha\neq 1$. The second term between square brackets comes from the cooling term: $-\int_0^r  n^2 \Lambda(T,Z) r'^2 dr'/\rho u r^2$.  Here, $n=\rho/\mu_{\rm n}$ is the number density of particles participating in the radiative process, $\mu_{\rm n}$ is the mean mass per particle for a neutral gas, and $\Lambda(T,Z)$ is the  radiative cooling function, which depends on both temperature and metallicity. For a given $Z$,  it is not expected to vary more than by a factor of $\sim 2$  in the interval of temperatures of interest  $\sim 10^6$--$10^8$ K, and thus, for simplicity,  it is taken to be $\Lambda=\Lambda(T_{\rm c},Z)$, where $T_{\rm c}$ is the temperature\footnote{Since radiative looses are expected to be larger at the centre of the object.} at $r=0$ (Section \ref{sol}).
Using the boundary condition at $R=1$, $U_{\rm sc}=C_{\rm sc}$,  we obtain  a quadratic equation for $U_{\rm sc}$, with roots:
\[U_{\rm sc}=\frac{1}{2(2\eta+1)} \left\{\left[{1-\frac{V_{\rm e}(1+\frac{1}{A})}{(5-2\alpha)}}{}\right] \; \pm\right.\]

\begin{equation}
\left. \sqrt{ \left[{1-\frac{V_{\rm e}(1+\frac{1}{A})}{(5-2\alpha)}}{}\right]^2-\frac{16(2\eta+1)\dot E \Lambda \xi^2 f(1)}{4\pi \mu_{\rm n}^2 r_{\rm sc} V_{\infty}^6 } \frac{(1+\frac{1}{A})}{(5-2\alpha)} }\,\right\}.\label{radth}
\end{equation}

From the above formula, we deduce that at least to the leading order, the gravitational inhibition of the flow  is uncoupled  from and is more decisive (i.e. it is stronger or faster) than  the inhibition by radiative cooling, since when the quantity between square brackets vanishes no stationary solution exist at all, independently of the cooling rate.  In that case, the flow adopts  one of the non-stationary  regimes previously described,  according to the value of $V_{\rm e}$. On the other hand, radiative cooling operates only when the gravitational field has not inhibited the solution and inhibits the flow when the quantity under the radical symbol vanishes or becomes negative.  From this condition, it is extremely easy to obtain the threshold (maximum) energy deposition rate, $\dot E_{\rm th}$,  for which cooling inhibits the stationary solution in terms of $V_{\infty}$, $r_{\rm sc}$, $\alpha$, $A$ and $V_{\rm e}$. The last three parameters are the new ingredients incorporated in our formulation. When  $V_{\rm e}=0$, $\alpha=0$ and $A\rightarrow \infty$ we recover the radiative threshold lines for the standard CC85 model estimated first numerically by Silich, Tenorio-Tagle \& Rodr\'iguez-Gonz\'alez (2004), given analytically by A\~norve-Zeferino (2006) using a necessary and sufficient condition for the existence of the stationary flow, and generalized later semi-analytically by W\"unsch et al. (2007) for CC85-like bimodal outflows. We are also able to reproduce as a particular case  the threshold lines  numerically obtained by Silich et al. (2010) for  a uniform distribution including the gravitational field ($V_{\rm e} \neq 0$, $\alpha=0$ and  $A\rightarrow \infty$).

Equation (\ref{radth}) contains additional useful information. The physically meaningful root corresponds to the '+' sign. This tells us that the radiative losses per unit volume cannot be more than $100(1 - 1/\sqrt{2})\%\sim 29\%$ of the effective energy injection rate \emph{minus} the rate at which work is done against the gravitational field, per unit volume. This is in outstanding agreement with previous numerical estimates   for the uniform distribution without gravitational field (Silich et al. 2004; W\"unsch et al. 2008; see, also, Strickland \& Heckman 2009), and thus, it  also confirms the validity of our leading order approach. This also indicates that as in the  previous numerical studies, when the stationary solution exist,
cooling is not going to modify significantly\footnote{N.B. In the purely radiative case, when cooling exceeds 29\%, it breaks up the stationary solution. In general, it is much less than that.} the  hydrodynamical profiles established by equations (\ref{CMN})--(\ref{CEN}).

In astrophysical units, the expression for $\dot E_{\rm th}$ is

\begin{equation}
\dot E_{\rm th}=(1.1\times 10^{45} \mbox{erg s$^{-1}$})  \frac{(5-2\alpha)r_{\rm sc,pc} V_{8}^6 \left[{1-\frac{V_{\rm e}(1+\frac{1}{A})}{(5-2\alpha)}}{}\right]^2 }{(2\eta+1)\left(1+1/A\right)\xi^2f_1\Lambda_1}, \label{Ethcool}
\end{equation}

\noindent where the radius is expressed in parsecs, the effective  terminal speed in units of 1000 km s$^{-1}$ and  $\Lambda_1$  is $\Lambda(T_{\rm c},Z)$ normalized to $10^{-23}$ erg  s$^{-1}$ cm$^3$.

\noindent The function $f_1=f(A,\alpha, 1)$ is given by

\[
f_1=\frac{(1+A)^{-2\alpha}}{A^2(\alpha-2)(2\alpha-3) } \left\{A^{2\alpha}(1+A)^5 \,- \right.
\]

\begin{equation}
{\left.A^3(1+A)^{2\alpha}\left[A^2+A(5-2\alpha)+(\alpha-2)(2\alpha-5)\right]\right\}}.
\end{equation}

Although cumbersome, the last function is needed to properly  locate the radiative threshold lines in the parameter space ($f$ came from direct integration). Nevertheless, a useful property is that $f(A,1,1)=1$, i.e. when evaluated at $R=1$, $f$ is independent of $A$ when $\alpha=1$. Other values of interest are  $f(0.5,0,1)=4$  and $f(0.1,0,1)=80/3$. This  interest rest on that we obtain asymptotically flat rotation curves for $\alpha\in[0,1]$ when $A\sim$ 0.1--0.5.

In turn, the weight factor is given by $\xi^2=b^2\xi_0^2 $, with
\begin{equation}
b=(2\eta+1)^{1/2} \left[\frac{b_0}{(2\eta+1)}\right]^{\frac{b_0}{2\left[\eta(1-\alpha)+(5-\alpha)(\eta+1)\right]}},
\end{equation}

\begin{equation}
b_0=(1-\alpha)(3\eta+1) + (5-\alpha)(\eta+1)
\end{equation}

\noindent and
\[ \xi_0^2= \exp\left(\frac{2\eta(\alpha+1)\alpha^2}{(A+1)^{5-\alpha}}+\frac{A}{A+1}\right)\times\] 
\begin{equation}
\hspace{6mm}\left(1+\frac{1}{(2\eta+1)A}\right)^{(1-(2\eta+1)A)}.
\end{equation}

\noindent The expression for $b$ recovers the exact value of the slope for velocity when $r\rightarrow 0$ and $A\rightarrow \infty$, and $\xi_0$ is a parametrization that accounts for the error in the approximation ($\xi_0\rightarrow 1$ when $A\rightarrow \infty$).

We will close the current discussion by  giving the expression of the galaxy radius for which a flow regime is inhibited by gravity for a given dynamical mass and effective terminal speed:

\begin{equation}
r_{\rm sc,grav}=\frac{2GM_{\rm DM}} {V_\infty^2V_{\rm e,th}},
\end{equation}

\noindent where $V_{\rm e,th}$ corresponds to the threshold $V_{\rm e}$ for one of the possible regimes: accelerating superwind, equation (\ref{Vecons});  bounded decelerating superwind or possible eventual retention with open-box enrichment,  equation (\ref{Vepouring}); or full retention with  possible open box enrichment, equation (\ref{Vecb}).

\end{document}